\documentclass[preprintnumbers,nofootinbib,noshowpacs,eqsecnum,prd,superscriptaddress,letterpaper]{revtex4}

\usepackage{graphicx}
\usepackage{amsmath,amssymb}
\usepackage{url}
\usepackage{multirow}
\usepackage{hyperref,color}

\newcommand{\sla}[1]{/\!\!\!\!#1}

\newcommand\one{\leavevmode\hbox{\small1\normalsize\kern-.33em1}}

\newcommand{\lag}{\mathcal{L}}

\newcommand{\ope}{\mathcal{O}}
\newcommand{\qqquad}{\qquad \qquad}
\newcommand{\qqqquad}{\qquad \qquad \qquad}

\newcommand{\gev}{\text{GeV}}
\newcommand{\tev}{\text{TeV}}

\newcommand{\ifb}{\text{fb}^{-1}}

\def\slashchar#1{\setbox0=\hbox{$#1$}           
   \dimen0=\wd0                                 
   \setbox1=\hbox{/} \dimen1=\wd1               
   \ifdim\dimen0>\dimen1                        
      \rlap{\hbox to \dimen0{\hfil/\hfil}}      
      #1                                        
   \else                                        
      \rlap{\hbox to \dimen1{\hfil$#1$\hfil}}   
      /                                         
   \fi}

\newcommand{\eg}{\textsl{e.g.}\;}

\DeclareMathOperator{\tr}{Tr}

\newcommand{\be}{\begin{eqnarray*}}
\newcommand{\ee}{\end{eqnarray*}}

\newcommand{\bee}{\begin{eqnarray}}
\newcommand{\eee}{\end{eqnarray}}
\newcommand{\beeq}{\begin{equation}}
\newcommand{\eeeq}{\end{equation}}

\begin{document}

\title{The Gauge-Higgs Legacy of the LHC Run I}

\preprint{YITP-SB-16-10} 

\author{Anja Butter}
\affiliation{Institut f\"ur Theoretische Physik, Universit\"at Heidelberg, Germany}
\author{Oscar J. P. \'Eboli}
\affiliation{Instituto de F\'isica, Universidade de S\~ao Paulo, S\~ao Paulo, Brazil}
\author{J.~Gonzalez--Fraile}
\affiliation{Institut f\"ur Theoretische Physik, Universit\"at Heidelberg, Germany}
\author{M.~C.~Gonzalez--Garcia}
\affiliation{C.N. Yang Institute for Theoretical Physics, SUNY at Stony Brook, USA}
\affiliation{Departament d'Estructura i Constituents de la Mat\`eria and ICC-UB Universitat de Barcelona, Spain}
\affiliation{Instituci\'o Catalana de Recerca i Estudis Avancats (ICREA)}
\author{Tilman Plehn}
\affiliation{Institut f\"ur Theoretische Physik, Universit\"at Heidelberg, Germany}
\author{Michael Rauch}
\affiliation{Institute for Theoretical Physics, Karlsruhe Institute of Technology, Germany}

\begin{abstract}
  The effective Lagrangian expansion provides a framework to study
  effects of new physics at the electroweak scale. To make full use of
  LHC data in constraining higher-dimensional operators we need to
  include both the Higgs and the electroweak gauge sector in our
  study. We first present an analysis of the relevant di-boson
  production LHC results to update constraints on triple gauge boson
  couplings. Our bounds are several times stronger than those obtained
  from LEP data.  Next, we show how in combination with Higgs
  measurements the triple gauge vertices lead to a significant
  improvement in the entire set of operators, including operators
  describing Higgs couplings.
\end{abstract}

\maketitle
\tableofcontents
\newpage

\section{Introduction}
\label{sec:intro}

The direct exploration of the electroweak symmetry breaking sector
started with the discovery of a light narrow Higgs boson~\cite{higgs}
in 2012~\cite{discovery} --- a triumph of particle physics. Already
the LHC Run~I allowed ATLAS and CMS to perform a large number of tests
of the nature of the observed resonance, but no significant deviations
from the Standard Model properties were observed for example in the
Higgs production and decay rates~\cite{legacy, other_final_fits,
  other_final_fits2, Ellis:2014jta}. On the other hand, it is
important to remind ourselves that the current constraints are still
at a precision level for which no significant deviations would be
expected in weakly interacting models of new
physics~\cite{bsm_review}. \medskip

If we accept the Standard Model assumption that the Higgs particle is
closely related to the massive gauge bosons, the Higgs results from
Run~I should be combined with corresponding precision measurements in
the electroweak sector. During Run~I the LHC collaborations have
also collected meaningful event samples probing electroweak gauge boson pair
production. They contain information on the structure of the
triple gauge boson vertices (TGV)s and allow for complementary tests
of the electroweak symmetry breaking mechanism. \medskip

The eventual observation of departures of Higgs or gauge boson
couplings from their SM predictions can give hints of physics beyond
the Standard Model, affecting the electroweak sector and characterized
by a new energy scale $\Lambda$. One way of parametrizing low--energy
effects of SM extensions is by means of an effective
Lagrangian~\cite{effective}, which only depends on the low--energy
particle content and symmetries. This bottom--up approach has the
advantage of minimizing the amount of theoretical hypothesis when
studying deviations from the SM predictions. Here, we assume that the
observed Higgs--like state belongs to a light electroweak doublet and
that the $SU(2)_L \otimes U(1)_Y$ symmetry is linearly realized in the
effective theory~\cite{effective-linear, kilian, higgsmultiplets,
  kaoru, Grzadkowski:2010es}.  Without lepton number violation, the
lowest order operators which can be built are of dimension six. The
coefficients of these dimension--six operators parametrize our
ignorance of the new physics effects and have to be determined using
all available data.\medskip

One result of this effective theory approach is that modified Higgs
couplings to weak bosons are related to triple gauge boson vertices in
a model independent fashion. This allows us to use Higgs data not only
to constrain TGVs~\cite{barca2}, but also to use TGV data to test the
strengths and structures of Higgs couplings. Usually, such combined
analyses rely on LEP results for the TGVs~\cite{barca, Brivio:2013pma,
  Falkowski:2015jaa}, the only exception being
Ref.~\cite{Ellis:2014jta}. The reason is that LEP provided the
strongest constraints on TGVs until now. However, during the LHC Run~I
both ATLAS and CMS have collected a substantial amount of data on
di-boson searches. It contains information on TGVs, whose relevance
has not been addressed quantitatively. We fill this void with the
first global analysis of the complete di-boson and Higgs data from the
LHC Run~I.\medskip

The outline of the paper is as follows: after briefly reviewing the
relevant set of operators in Sec.~\ref{sec:intro_th}, we present the
results of our global analysis of the LHC Run~I data on di-boson
searches in Sec.~\ref{sec:gauge}.  We find that the combined LHC Run~I
results are substantially stronger than the LEP constraints.
Section~\ref{sec:higgs} contains the combined analysis of di-boson and
Higgs data, giving the up-to-date limits on the ten relevant Wilson
coefficients.  We summarize in Sec.~\ref{sec:conclu}. The details of
our di-boson simulations can be found in the Appendix~\ref{sec:app}. 

\section{Theoretical framework}
\label{sec:intro_th}

In the linear effective Lagrangian expansion we construct a
$SU(3)_c \otimes SU(2)_L \otimes U(1)_Y$-symmetric Lagrangian based on
the SM field content, including the Higgs-Goldstone doublet $\phi$. We
order the Lagrangian according to the inverse powers of the new
physics scale~\cite{effective-linear, kilian, higgsmultiplets, kaoru,
  Grzadkowski:2010es},
\begin{alignat}{5}
\lag = \sum_x \frac{f_x}{\Lambda^2} \; \ope_x \;\;,
\label{eq:def_f}
\end{alignat}
where $\Lambda$ is the natural choice for a matching scale with a
given complete theory.  Neglecting the dimension--five lepton number
violating operator the next order of the expansion is based on
dimension--six operators. \medskip

The minimum independent set consists of 59 baryon number conserving
operators, barring flavor structure and Hermitian
conjugation~\cite{Grzadkowski:2010es}. We follow the definition of the
relevant operator basis for LHC Higgs and TGV physics described in
detail in Ref.~\cite{barca}. We start by restricting the initial set
to $P$ and $C$--even operators. We then use the equations of motion to
rotate to a basis where there are no blind directions linked to
electroweak precision data. In practice, we can neglect all operators
contributing to electroweak precision observables at tree level; they
are strongly constrained by the several low energy measurements,
rendering them irrelevant for current Higgs and TGV studies at the
LHC.  We then neglect all operators that cannot be studied at the LHC
yet, because they only contribute to interactions we are not sensitive
to. This includes the $HHH$ vertex or the Higgs interactions with
light-generation fermions.  Finally, we are left with ten
dimension--six operators~\cite{barca}:
\begin{alignat}{9}
\ope_{GG} &= \phi^\dagger \phi \; G_{\mu\nu}^a G^{a\mu\nu}  \qqqquad 
&\ope_{WW} &= \phi^{\dagger} \hat{W}_{\mu \nu} \hat{W}^{\mu \nu} \phi  \qqqquad 
&\ope_{BB} &= \phi^{\dagger} \hat{B}_{\mu \nu} \hat{B}^{\mu \nu} \phi 
\notag \\
\ope_W &= (D_{\mu} \phi)^{\dagger}  \hat{W}^{\mu \nu}  (D_{\nu} \phi)
& \ope_B &=  (D_{\mu} \phi)^{\dagger}  \hat{B}^{\mu \nu}  (D_{\nu} \phi) 
&\ope_{\phi,2} &= \frac{1}{2} \partial^\mu\left ( \phi^\dagger \phi \right)
                            \partial_\mu\left ( \phi^\dagger \phi \right) \qquad
\notag \\
\ope_{e\phi,33} &=(\phi^\dagger\phi)(\bar L_3 \phi e_{R,3}) 
\qquad 
&\ope_{u\phi,33} &=(\phi^\dagger\phi)(\bar Q_3 \tilde \phi u_{R,3})
\qquad  
&\ope_{d\phi,33} &=(\phi^\dagger\phi)(\bar Q_3 \phi d_{R,3}) \; \notag \\
\ope_{WWW} &= \tr \left( \hat{W}_{\mu \nu} \hat{W}^{\nu \rho} 
\hat{W}_\rho^\mu \right)  \; . 
\label{eq:operators}  
\end{alignat}
In our conventions the Higgs doublet covariant derivative is
$D_\mu\phi= \left(\partial_\mu+ i g' B_\mu/2 + i g \sigma_a W^a_\mu/2
\right)\phi $.
The hatted field strengths are
$\hat{B}_{\mu \nu} = i g' B_{\mu \nu}/2$ and
$\hat{W}_{\mu\nu} = i g\sigma^a W^a_{\mu\nu}/2$, where $\sigma^a$ are
the Pauli matrices, and $g$ and $g^\prime$ stand for the $SU(2)_L$ and
$U(1)_Y$ gauge couplings. The adjoint Higgs field is
$\tilde \phi=i\sigma_2\phi^*$. The effective Lagrangian which we use
to interpret Higgs and TGV measurements at the LHC is
\begin{align}
\lag_\text{eff} = \lag_\text{SM}
& - \frac{\alpha_s }{8 \pi} \frac{f_{GG}}{\Lambda^2} \ope_{GG}  
+ \frac{f_{BB}}{\Lambda^2} \ope_{BB} 
+ \frac{f_{WW}}{\Lambda^2} \ope_{WW} + \frac{f_{\phi,2}}{\Lambda^2} \ope_{\phi,2}
+ \frac{f_{WWW}}{\Lambda^2} \ope_{WWW} \nonumber\\
&+ \frac{f_B}{\Lambda^2} \ope_B  + \frac{f_W}{\Lambda^2} \ope_W
+ \frac{f_\tau m_\tau}{v \Lambda^2} \ope_{e\phi,33} 
+ \frac{f_b m_b}{v \Lambda^2} \ope_{d\phi,33} 
+ \frac{f_t m_t}{v \Lambda^2} \ope_{u\phi,33}\;.
\label{eq:ourlag}
\end{align}
All operators except for $\ope_{WWW}$ contribute to Higgs
interactions.  Their contributions to the several Higgs vertices,
including non--SM Lorentz structures, are described in
Ref.~\cite{legacy}.\medskip
 
Some of the operators in Eq.~\eqref{eq:operators} contribute to the
self-interactions of the electroweak gauge bosons. They can be linked
to specific deviations in the Lorentz structures entering the $WWZ$
and $WW\gamma$ interactions, usually written as
$\kappa_\gamma, \kappa_Z, g_1^Z, g_1^\gamma$, $\lambda_\gamma$, and
$\lambda_Z$~\cite{Hagiwara:1986vm}. After $g_1^\gamma$ is fixed to
unity because of electromagnetic gauge invariance, writing the
deviations with respect to the SM values for example as
$\Delta \kappa \equiv \kappa-1$, the shifts are defined as
\begin{align}
\Delta \lag_\text{TGV} =& 
- i e \; \Delta \kappa_\gamma \; W^+_\mu W^-_\nu \gamma^{\mu \nu}
- \frac{i e \lambda_\gamma}{2 m_W^2} \; W_{\mu \nu}^+ W^{- \nu \rho} \gamma_\rho^\mu
- \frac{i g_Z \lambda_Z}{2 m_W^2} \; W_{\mu \nu}^+ W^{- \nu \rho} Z_\rho^{\;\mu} 
\notag \\
&- i g_Z \; \Delta \kappa_Z \; W^+_\mu W^-_\nu Z^{\mu \nu}
- i g_Z \; \Delta g_1^Z \; \left( W^+_{\mu \nu} W^{- \mu} Z^\nu - W^+_\mu Z_\nu W^{- \mu \nu} 
                    \right) \notag \\
=& - i e \; \frac{g^2 v^2}{8 \Lambda^2} \left( f_W + f_B \right)  \; W^+_\mu W^-_\nu \gamma^{\mu \nu}
- i e \; \frac{3 g^2 f_{WWW}}{4 \Lambda^2} \; W_{\mu \nu}^+ W^{- \nu \rho} \gamma_\rho^\mu \notag \\
&- i g_Z \; \frac{g^2 v^2}{8 c_w^2 \Lambda^2} \left( c_w^2 f_W - s_w^2 f_B \right) \; W^+_\mu W^-_\nu Z^{\mu \nu}
- i g_Z \; \frac{3 g^2 f_{WWW}}{4 \Lambda^2} \; W_{\mu \nu}^+ W^{- \nu \rho} Z_\rho^{\; \mu} \notag \\
&- i g_Z \; \frac{g^2 v^2 f_W}{8 c_w^2 \Lambda^2} \; \left( W^+_{\mu \nu} W^{- \mu} Z^\nu - W^+_\mu Z_\nu W^{- \mu \nu} 
                    \right) \; ,
\label{eq:tgvlag}
\end{align}
where $e = g s_w$ and $g_Z = g c_w$. The two notational conventions are linked as
\begin{alignat}{9}
\Delta \kappa_\gamma &= 
 \frac{g^2 v^2}{8\Lambda^2}
\left( f_W + f_B \right) \qqqquad 
&\Delta \kappa_Z &=   \frac{g^2 v^2}{8 c_w^2\Lambda^2} \left(c_w^2 f_W - s_w^2 f_B \right) \notag \\  
\Delta g_1^Z &= \frac{g^2 v^2}{8 c_w^2\Lambda^2} f_W \qqqquad
&\Delta g_1^\gamma &= 0 
&\lambda_\gamma &= \lambda_Z = 
\frac{3 g^2 M_W^2}{2 \Lambda^2} f_{WWW}\; .
\label{eq:wwv}
\end{alignat}
The $SU(2)$-gauge-invariant formulation in terms of dimension--six
operators induces correlations of the formerly multi-dimensional space
of modified gauge couplings,
\begin{align}
\lambda_Z=\lambda_\gamma
\qquad \text{and} \qquad 
\Delta\kappa_Z =-\frac{s_w^2}{c_w^2}\Delta\kappa_\gamma +\Delta g_1^Z\;.
\label{eq:lep}
\end{align}
This defines what is usually referred to as the LEP scenario in the
analysis of anomalous TGV interactions. The three relevant Wilson
coefficients relevant for our analysis of di-boson production are
$f_B$, $f_W$ and $f_{WWW}$. 

\section{Triple gauge boson interactions}
\label{sec:gauge}

In our analysis we describe the measured di-boson production rates
from the LHC Run~I in terms of the Lagrangian given in
Eq.~\eqref{eq:tgvlag}.  We include the eight $WV (V = W,Z)$ di-boson
measurements with the highest sensitivity for charged triple gauge
boson vertices.  Adding the public $W\gamma$ LHC results, only
available for 7~TeV so far~\cite{cms7wa,atlas7wa}, does not improve
our results. \medskip

For each analysis we first determine which of the kinematic
distributions given in the publications is most sensitive to anomalous
TGVs. This defines our list of channels and kinematic variables, as
well as the available number of bins of the distribution.

\begin{center} \begin{tabular}{l|lcr}
\hline 
Channel & Distribution & \# bins  & Data set \\ [0mm]
\hline
$WW\rightarrow \ell^+\ell^{\prime -}+\sla{E}_T\; (0j)$ & Leading lepton $p_{T}$ & 4 & ATLAS 8 TeV, 20.3 fb$^{-1}$~\cite{atlas8ww}  \\[0mm]
$WW\rightarrow \ell^+\ell^{(\prime) -}+\sla{E}_T\; (0j)$ & $m_{\ell\ell^{(\prime)}}$ & 8 & CMS 8 TeV, 19.4 fb$^{-1}$~\cite{cms8ww}  \\[0mm]
$WZ\rightarrow \ell^+\ell^{-}\ell^{(\prime)\pm}$ & $m_{T}^{WZ}$ & 6 & ATLAS 8 TeV, 20.3 fb$^{-1}$~\cite{atlas8wz}  \\[0mm]
$WZ\rightarrow \ell^+\ell^{-}\ell^{(\prime)\pm}+\sla{E}_T$ & $Z$ candidate $p_{T}^{\ell\ell}$ & 10 & CMS 8 TeV, 19.6 fb$^{-1}$~\cite{cms78wz}  \\[0mm]
$WV\rightarrow \ell^\pm jj+\sla{E}_T$& $V$ candidate $p_{T}^{jj}$ & 12 & ATLAS 7 TeV, 4.6 fb$^{-1}$~\cite{atlas7semilep} \\[0mm]
$WV\rightarrow \ell^\pm jj+\sla{E}_T$& $V$ candidate $p_{T}^{jj}$ & 10 & CMS 7 TeV, 5.0 fb$^{-1}$~\cite{cms7semilep} \\[0mm]
$WZ\rightarrow \ell^+\ell^{-}\ell^{(\prime)\pm}+\sla{E}_T$ & $Z$ candidate $p_{T}^{\ell\ell}$ & 7 & ATLAS 7 TeV, 4.6 fb$^{-1}$~\cite{atlas7wz}  \\[0mm]
$WZ\rightarrow \ell^+\ell^{-}\ell^{(\prime)\pm}+\sla{E}_T$ & $Z$ candidate $p_{T}^{\ell\ell}$ & 8 & CMS 7 TeV, 4.9 fb$^{-1}$~\cite{cms78wz}  \\[0mm]
\hline
\end{tabular} \end{center}

\noindent
In the final states only $\ell^{(\prime)} = e, \mu$ are considered,
channels with $(0j)$ include a jet veto, and the two semileptonic
channels include a veto on a third hard jet. 

\subsubsection{Analysis framework}

Directly from the relevant experimental figure we read off the
background expectation (defined as all SM processes except for the
di-boson production channels), the expected contribution from $WV$
production in the Standard Model and the measured event number bin by
bin. The background rates we use directly from the experimental
analysis, without any need to modify them.  Next, we simulate SM $WV$
production in the fiducial region using
\textsc{MadGraph5}~\cite{madgraph} for the event generation,
\textsc{Pythia}~\cite{Sjostrand:2006za} for parton shower and
hadronization, and \textsc{Delphes}~\cite{delphes} for the
detector simulation. We compare these results to the experimental
predictions, defining a bin-by-bin correction factor. It accounts for
phase--space dependent corrections either from detector effects or
from higher order corrections~\cite{nloww}. These correction factors we apply to
our simulated $WV$ distributions in the presence of the anomalous
TGVs, based on an in-house \textsc{MadGraph5} implementation of the
operators constructed with
\textsc{FeynRules}~\cite{Christensen:2008py}.  In the Appendix we give
more details on this procedure for one of the leading experimental channels,
{\it i.e.} the leptonic ATLAS $WW$ production at 8
TeV~\cite{atlas8ww}. \medskip

We check this default procedure using an alternative setup where
instead of matching our SM $WV$ distributions bin-by-bin, we only
match our inclusive $WV$ rate prediction in the Standard Model in the
signal region. Both methods give consistent results for the combined
analysis.\medskip

The parameter determination relies on \textsc{SFitter}, for technical
details we refer to Refs.~\cite{sfitter_orig,sfitter_delta,legacy}.
We first construct Markov chains in the three-dimensional model space
of $f_W$, $f_B$ and $f_{WWW}$. Then we build the likelihood function
for the given data set and determine the part of parameter space
allowed at a given CL. In the construction of the likelihood we always
include Poisson-shaped statistical uncertainties for event numbers, a
Gaussian-shaped experimental systematic uncertainty and a flat theory
uncertainty for the signal. As experimental systematics we include the
biggest sources of uncertainties for a given experiment, this includes
the luminosity estimate, detector and lepton reconstruction/isolation
uncertainties, and some additional uncertainty for the background
normalization and/or shape, all discussed in the Appendix. For the
theoretical uncertainty we allow for a variation of 5\% for $WW$, 4\%
for $WZ$ and 4\% for $WV$-semileptonic channels. We fully correlate
theoretical uncertainties for sets with the same di-boson final
state. \medskip

Wherever the experimental collaborations present their results in
terms of anomalous TGVs we validate our procedure through a detailed
comparison with their results as exemplified in the
Appendix~\ref{sec:app}. 

\subsubsection{Results from LHC Run~I}

\begin{figure}[t]
\includegraphics[width=0.32\textwidth]{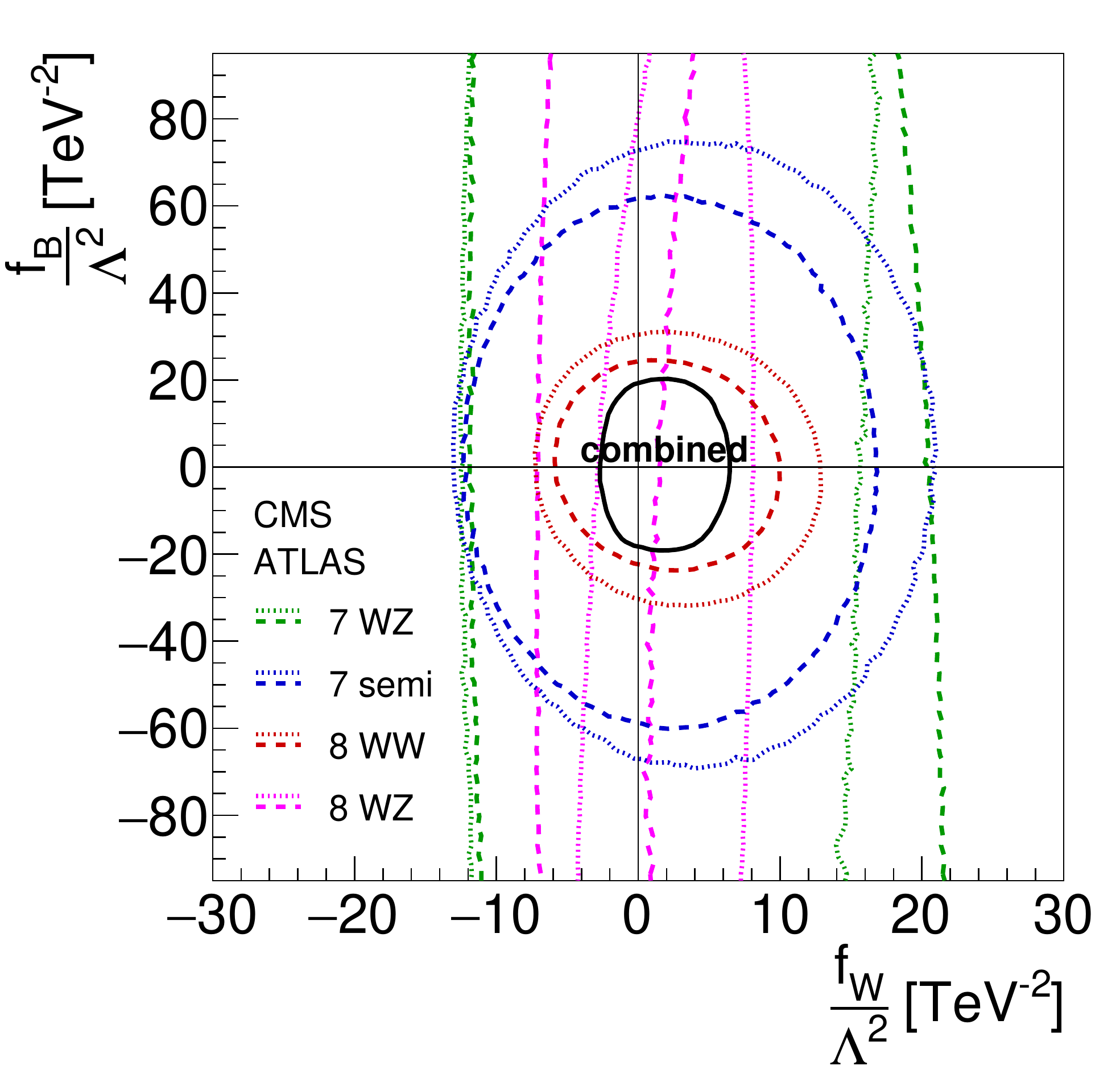}
\includegraphics[width=0.32\textwidth]{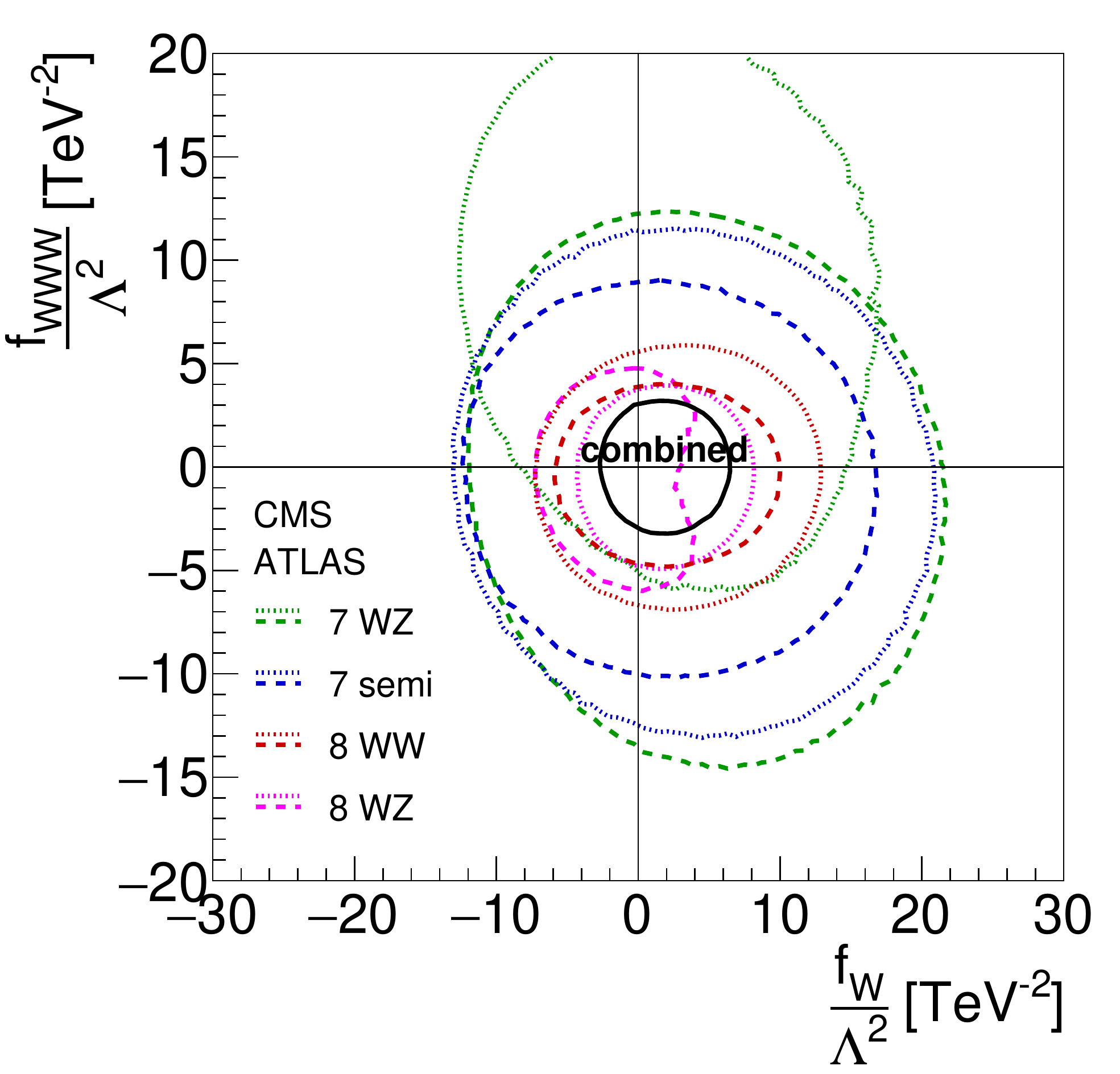}
\includegraphics[width=0.32\textwidth]{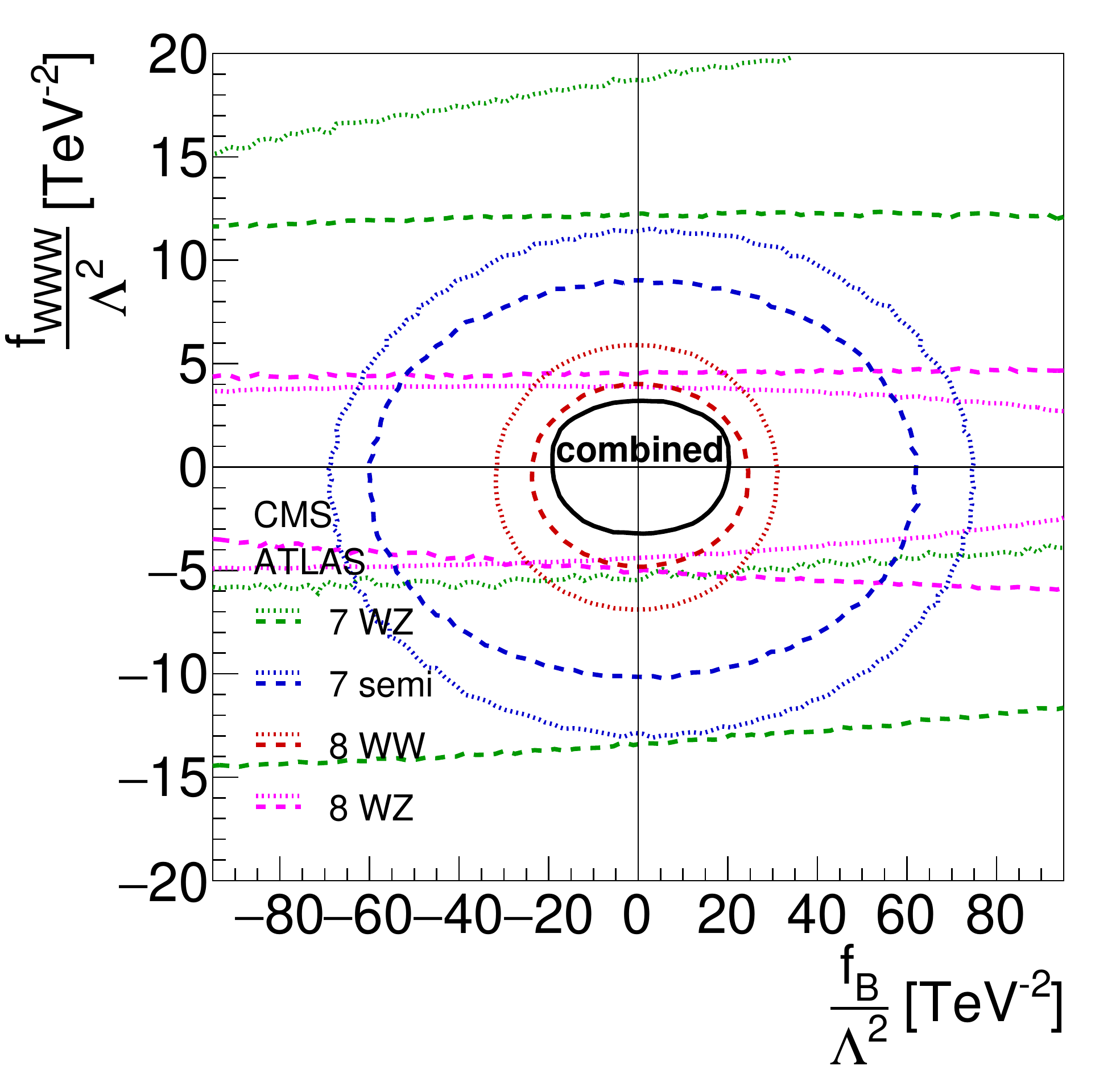}
\caption{Results of the TGV analysis from LHC Run~I. We show all
  two-dimensional profile likelihoods in the three-dimensional
  parameter space at 95\% CL (2dof) for the individual channels as
  well as their combination.}
\label{fig:tgv95CL}
\end{figure}

In Fig.~\ref{fig:tgv95CL} we show the results of our pure TGV analysis
in terms of the Wilson coefficients defined in Eq.~\eqref{eq:ourlag}.
In addition to each individual ATLAS and CMS channel we give the
combined constraints from all eight channels. For the combination, we
find a global minimum at a Gauss-equivalent
$\chi^2 \approx - 2 \log L = 48.3$ for a total of 65 data points,
while $\chi^2 \approx - 2 \log L =49.7$ for the Standard Model. The
regions allowed by the different searches are mutually compatible and
show no significant preference for a deviation from the Standard
Model. Moreover, the structure of the parameter space is simple enough
that none of the two-dimensional planes significantly change if
instead of a profile likelihood we show a slice where the third Wilson
coefficient is zero.\medskip

The Wilson coefficient $f_B$ is the least constrained because it
hardly affects the $WWZ$ vertex since its contribution is suppressed
by a factor $s_w^2/c_w^2$.  Instead, the constraints on $f_B$ come
from the fully leptonic $WW$ searches and to some degree from the
$WV$-semileptonic analyses, both probing the $WW\gamma$ interaction.
The ATLAS $WW$ channel at 8~TeV sets the strongest bounds on
$f_B$. \medskip

Comparing $f_W$ and $f_{WWW}$, we notice that the combination of the
$WWZ$ and $WW\gamma$ vertices with the large transverse momentum
available at the LHC leads to similar sensitivities on both;
equivalently, we find comparable sensitivities on $\lambda_{\gamma,Z}$
and $\Delta g_1^Z$. The new physics reach in $f_W$ and $f_{WWW}$ is
clearly stronger than in $f_B$. The strongest bounds on $f_{WWW}$ stem
from the combination of the two 8~TeV $WZ$ leptonic searches together
with the ATLAS 8~TeV $WW$ analysis. In the case of $f_W$, the 8~TeV
$WZ$ analyses present a higher sensitivity, but again the 8~TeV $WW$
searches are close in their precision. The constraint on $f_W$
benefits most from a combination of the different experimental
channels. \medskip

Generally, even though the $WV$-semileptonic results presented here
are less sensitive to the dimension--six operators, they are not far
from the most powerful leptonic $WW$ and $WZ$ analyses. This is
remarkable, given the fact that these semileptonic measurements are
still based on the 7~TeV smaller data sets. An update of the
semileptonic channels should significantly contribute to a global TGV
analysis.\medskip

The one-dimensional 95\%~CL constraints on the combination of Wilson
coefficients are
\begin{align}
\frac{f_W}{\Lambda^2} &\in \left[-1.5,\ 6.3\,\right]~\tev^{-2} \qquad 
&\frac{f_B}{\Lambda^2} &\in \left[-14.3,\ 15.9\,\right]~\tev^{-2} \qquad 
&\frac{f_{WWW}}{\Lambda^2} &\in \left[-2.4,\ 3.2\,\right]~\tev^{-2} \; .
\label{eq:tgvLHCI95ranges}
\end{align}
The same results can also be expressed as
\begin{align}
\frac{\Lambda}{\sqrt{|f_W|}} &> 0.82 \left( 0.40\right)~\tev \qquad 
&\frac{\Lambda}{\sqrt{|f_B|}} &> 0.26 \left( 0.25\right)~\tev \qquad 
&\frac{\Lambda}{\sqrt{|f_{WWW}|}} &> 0.65 \left( 0.56\right)~\tev \; ,
\end{align}
where the bounds stand for the limits obtained assuming a negative
(positive) Wilson coefficient.  Moreover, we can present our results
in terms of three independent TGV couplings~\cite{Hagiwara:1986vm}, as
described in Sec.~\ref{sec:intro_th}, the 95\%~CL constraints then read
\begin{align}
\Delta g_1^Z\in \left[-0.006,\ 0.026\,\right] \qqqquad 
\Delta \kappa_\gamma \in \left[-0.041,\ 0.072\,\right] \qqqquad 
\lambda_{\gamma,Z} \in \left[-0.0098,\ 0.013\,\right]\; .
\label{eq:tgvLHCI95ranges_2}
\end{align}
One aspect that we have tested is how robust our results are when we
change our approximate treatment of fully correlated theoretical
uncertainties. It turns out that removing these correlations slightly
shifts the $f_W$ range towards negative values and weaken the bound on
$f_B$; both effects are at the level of less than 0.5 standard
deviations. \medskip

To allow for an easy presentation of the approximate fit results we
perform a Gaussian fit to the multi-dimensional probability
distribution function of the three Wilson coefficients relevant for
TGVs.  For the mean, one standard deviation and the error correlation
matrix we find
\begin{align}
\frac{f_W}{\Lambda^2} =\left( 2.2\pm 1.9 \right)~\tev^{-2} \qqquad 
&\frac{f_B}{\Lambda^2}=\left( 3.0\pm 8.4 \right)~\tev^{-2} \qqquad
\frac{f_{WWW}}{\Lambda^2}=\left( 0.55\pm 1.4 \right)~\tev^{-2} \notag \\
\rho&=\begin{pmatrix}
1.00 & -0.012 & -0.062 \\[-1mm]
-0.012 & 1.00 & -0.0012 \\[-1mm]
-0.062 & -0.0012 &1.00
\end{pmatrix} \; .
\label{eq:cormat}
\end{align}
The corresponding Gaussian fit results to the multi-dimensional probability
distribution function for the TGV couplings in Eq.~\ref{eq:wwv} are shown in
Table~\ref{tab:complep}. 

\subsubsection{Comparison and combination with LEP}

\begin{figure}[t]
\includegraphics[width=0.32\textwidth]{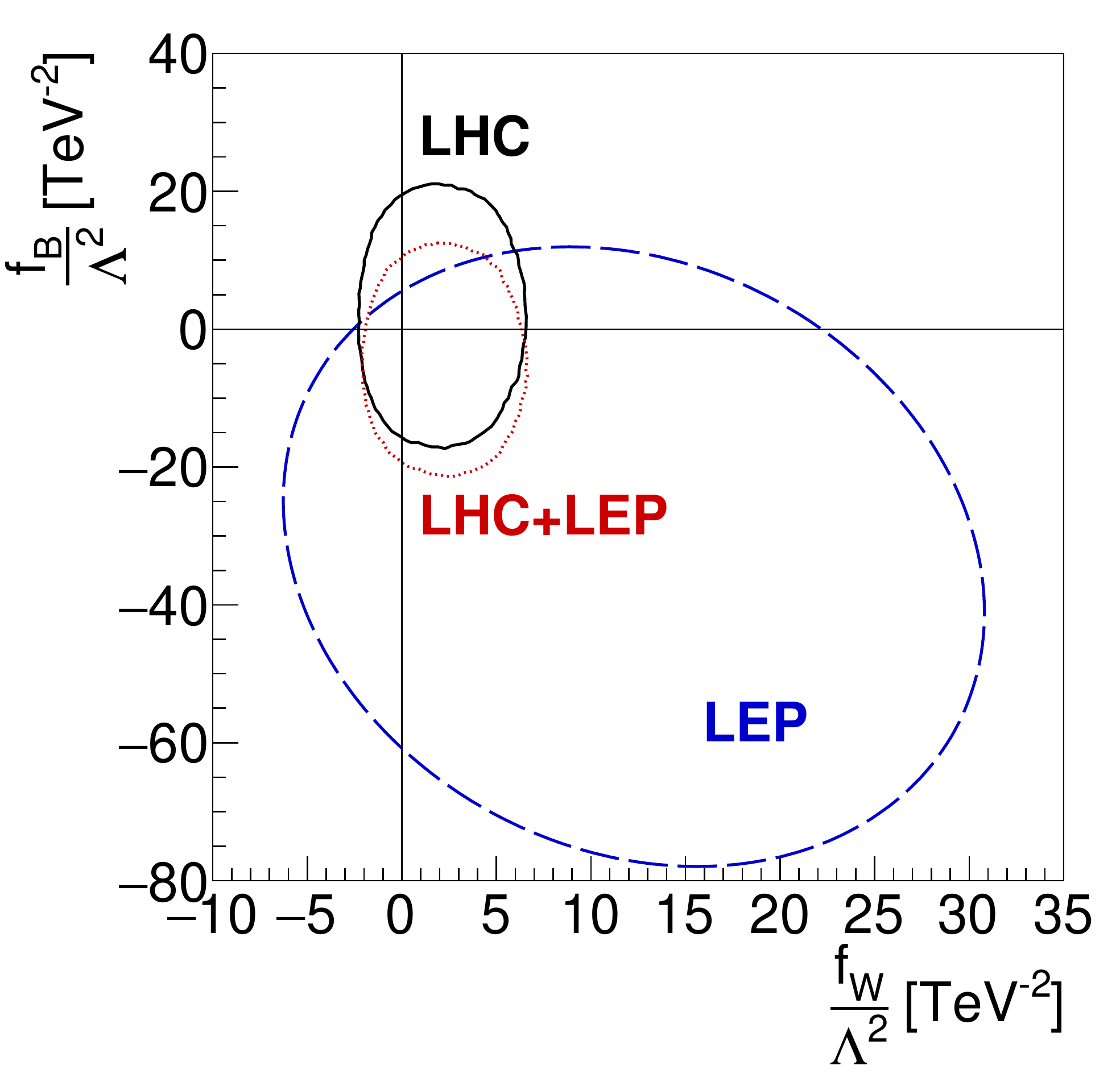}
\includegraphics[width=0.32\textwidth]{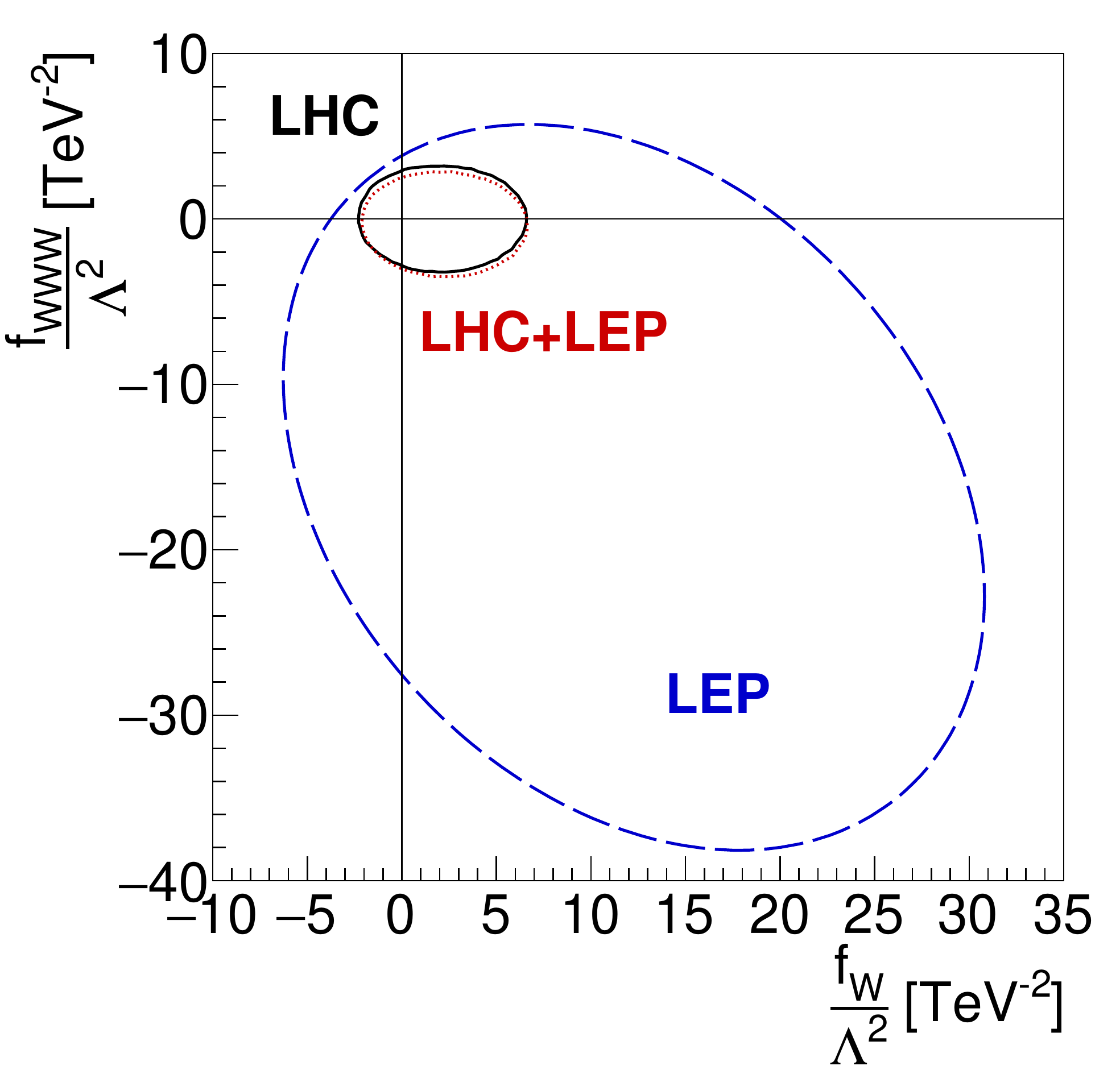}
\includegraphics[width=0.32\textwidth]{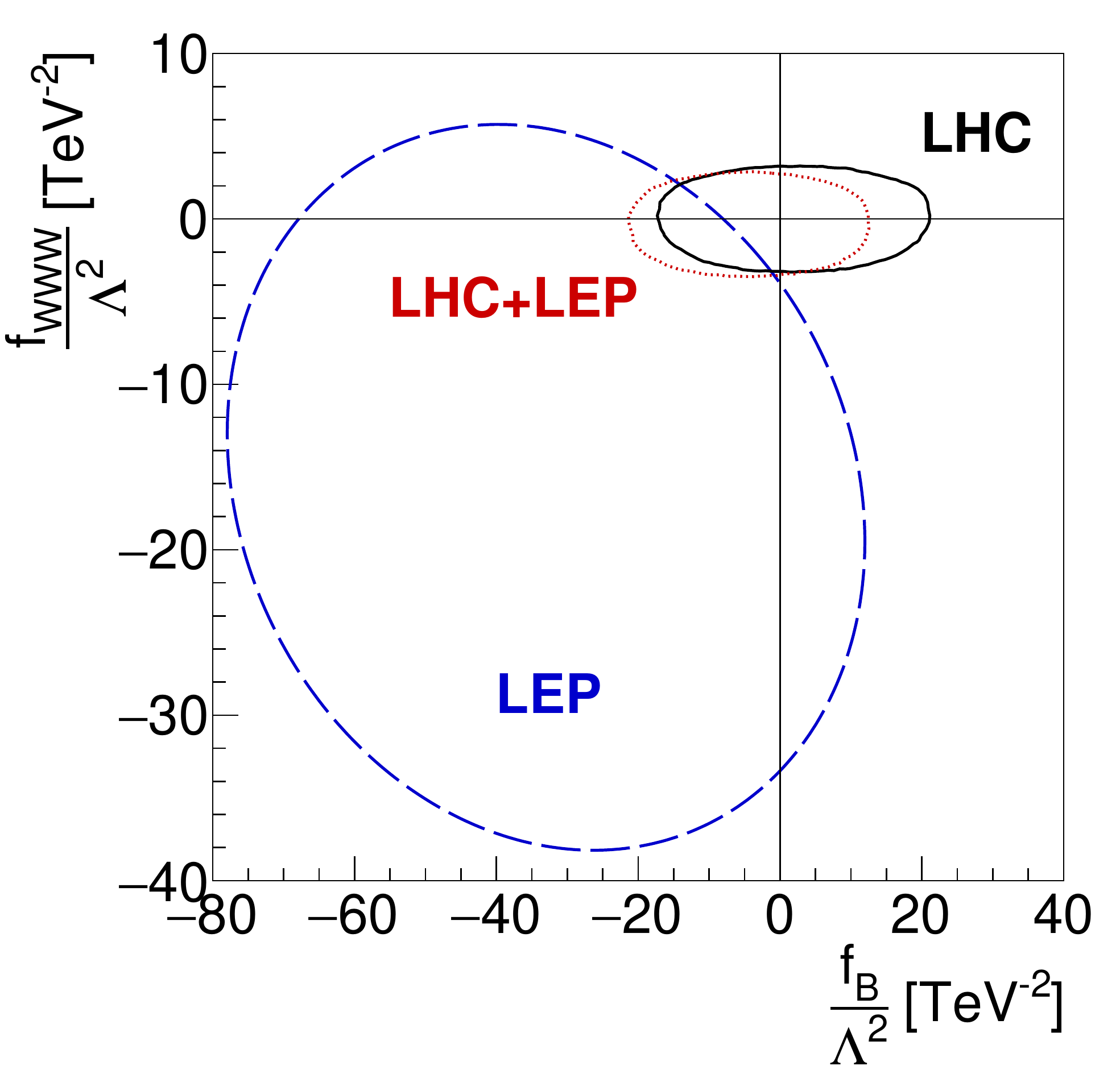}
\caption{Results of the TGV analysis in terms of two-dimensional
  profile likelihoods from LHC Run~I and from LEP~\cite{LEP2002}. We
  also show the statistical combination of both.}
\label{fig:tgvlep}
\end{figure}

When we express our results in terms of the TGVs defined in
Eq.~\eqref{eq:wwv} we can easily compare them and eventually combine
them with the global LEP analysis results~\cite{LEP2002}. We show the
separated LHC Run~I and LEP limits in Table~\ref{tab:complep}. As we
can see, the combined LHC Run~I di-boson channels determine the
anomalous TGV parameters a factor 3-6 more precisely than
LEP. Moreover, the more diverse set of LHC observables implies that
the different coupling measurements are less correlated.

\begin{table}[t]
\begin{tabular}{c|crrr|crrr}
\hline 
&\multicolumn{4}{c|}{ LHC Run I} &\multicolumn{4}{c}{LEP} \\ 
& \phantom{xxxxxx} 68 \% CL \phantom{xxxxxx} & \multicolumn{3}{c|}{Correlations} 
& \phantom{xxxxxx} 68 \% CL \phantom{xxxxxx} & \multicolumn{3}{c}{Correlations} \\ \hline
$\Delta g_1^Z$ & $0.010\pm 0.008$ & $1.00$ & $0.19$ & $-0.06$ 
               & \phantom{$-$} $0.051^{+0.031}_{-0.032}$ & $1.00$ & $0.23$ & $-0.30$ \\[1mm]
$\Delta \kappa_\gamma$ & $0.017\pm 0.028$ & $0.19$ & $1.00$ & $-0.01$ 
                      & $-0.067^{+0.061}_{-0.057}$ & $0.23$ & $1.00$ & $-0.27$ \\[1mm]
$\lambda$ & $0.0029\pm 0.0057$ & $-0.06$ & $-0.01$ & $1.00$ 
          & $-0.067^{+0.036}_{-0.038}$ & $-0.30$ & $0.27$ & $1.00$ \\
\hline
\end{tabular}
\caption{Measured central values, standard deviations and correlation
  coefficients for $\Delta g_1^Z$, $\Delta \kappa_\gamma$ and  $\lambda$
  from the combined LHC Run~I di-boson analyses (left) and from
  LEP~\cite{LEP2002} (right).}
\label{tab:complep}
\end{table}

The same comparison between the combined LHC Run~I results and the LEP
bounds is illustrated in Fig.~\ref{fig:tgvlep}, now in terms of
dimension-six Wilson coefficients. In these two-dimensional profile
likelihoods we also show the statistical combination of the two data
sets.  While the LHC precision shown in Eq.~\eqref{eq:tgvLHCI95ranges}
clearly dominates the combination of LHC and LEP results, we still
quote the combined limits on the three relevant Wilson coefficients,
\begin{align}
\frac{f_W}{\Lambda^2}\in \left[-1.3,\ 6.3\,\right]~\tev^{-2} \qqquad 
\frac{f_B}{\Lambda^2}\in \left[-18,5,\ 10.9\,\right]~\tev^{-2} \qqquad 
\frac{f_{WWW}}{\Lambda^2}\in \left[-2.7,\ 2.8\,\right]~\tev^{-2} \; .
\label{eq:tgvLHCLEPI95ranges}
\end{align}
Adding the LEP results does not lead to a significant improvement. The
range for $f_B$ slightly shifts towards more negative values as a
consequence of the preferred LEP central values. 

\section{Gauge--Higgs combination}
\label{sec:higgs}

\begin{figure}[b!]
  \includegraphics[width=0.3\textwidth]{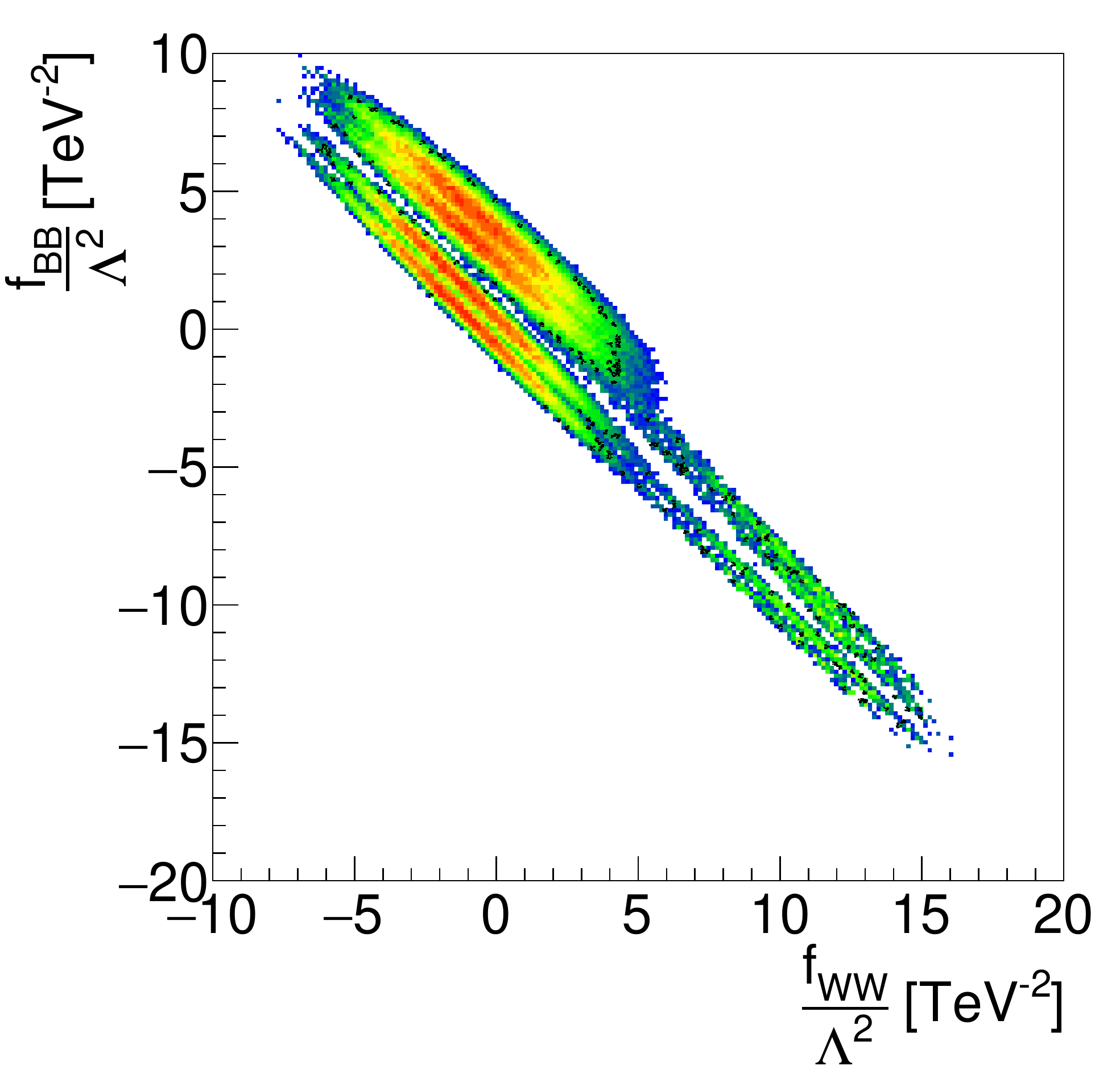}
  \hspace*{-0.2cm}
  \includegraphics[width=0.3\textwidth]{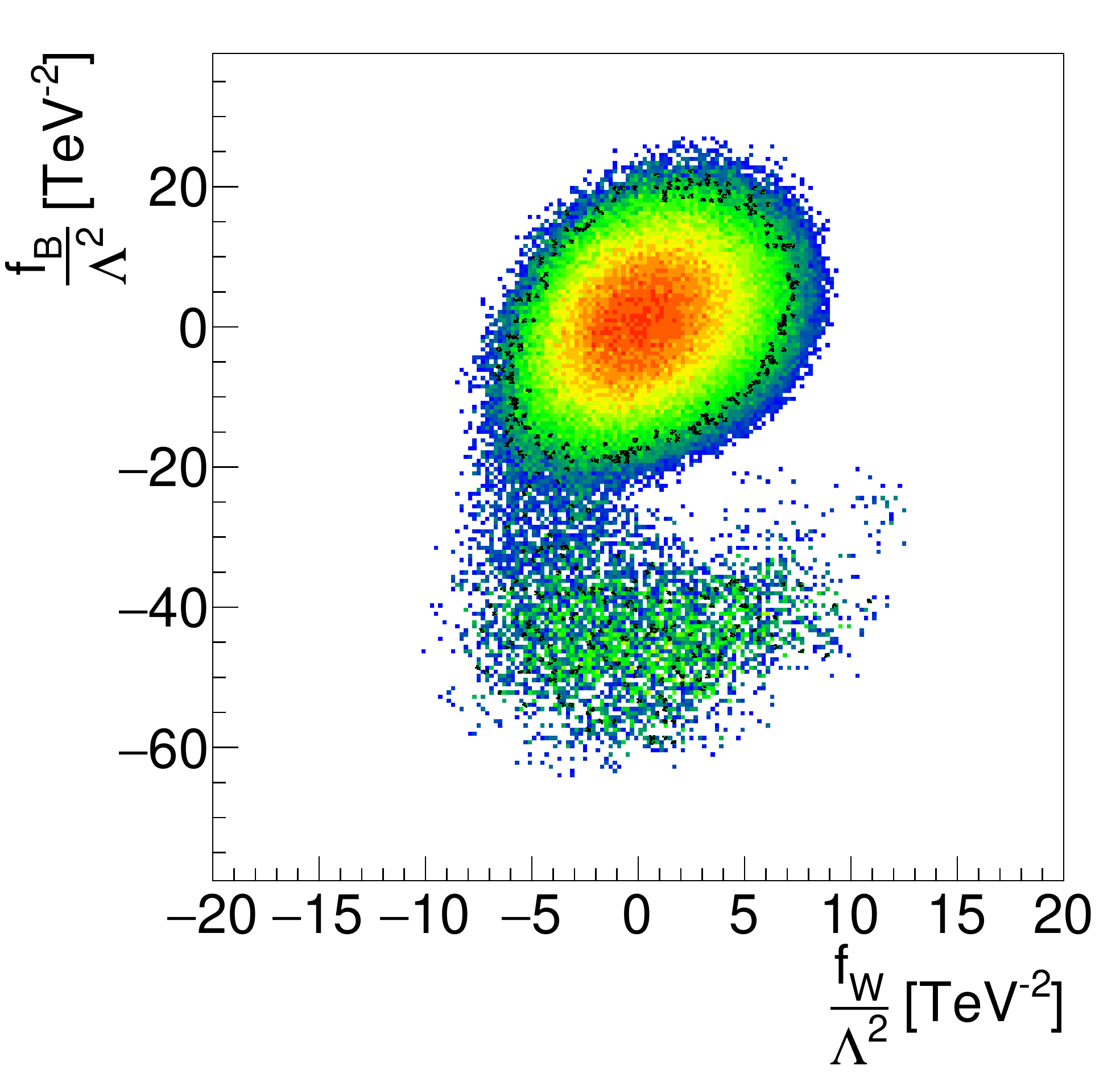}
  \hspace*{-0.2cm}
  \includegraphics[width=0.3\textwidth]{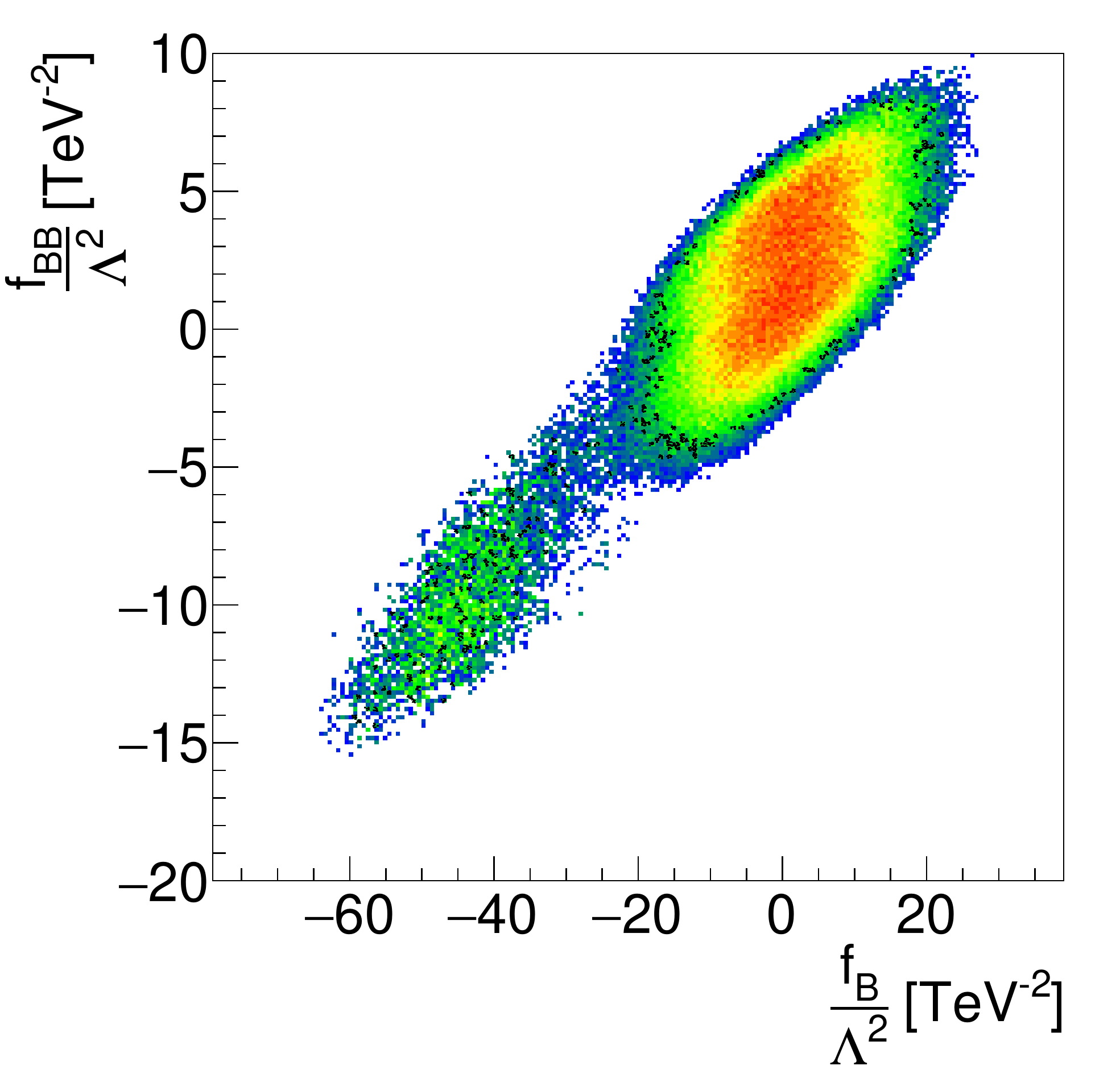}
   \raisebox{3pt}{\includegraphics[width=0.051\textwidth]{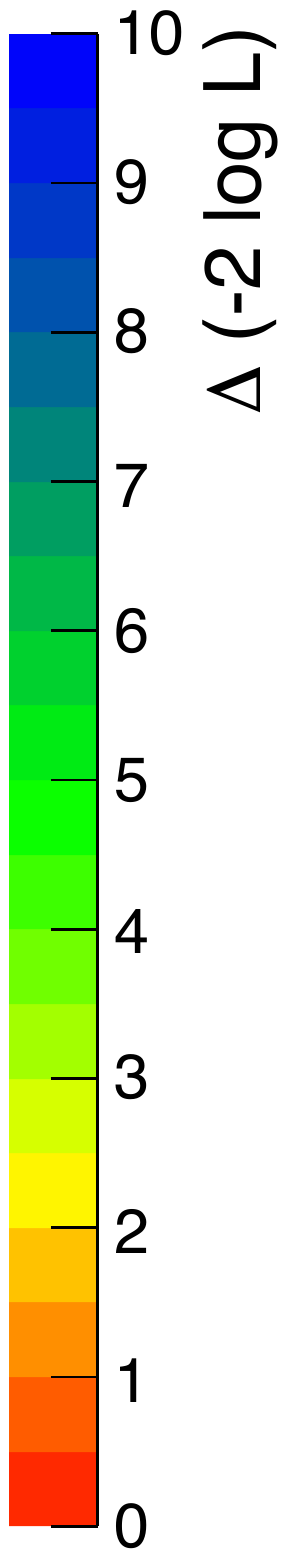}}\\[1ex]
  \includegraphics[width=0.3\textwidth]{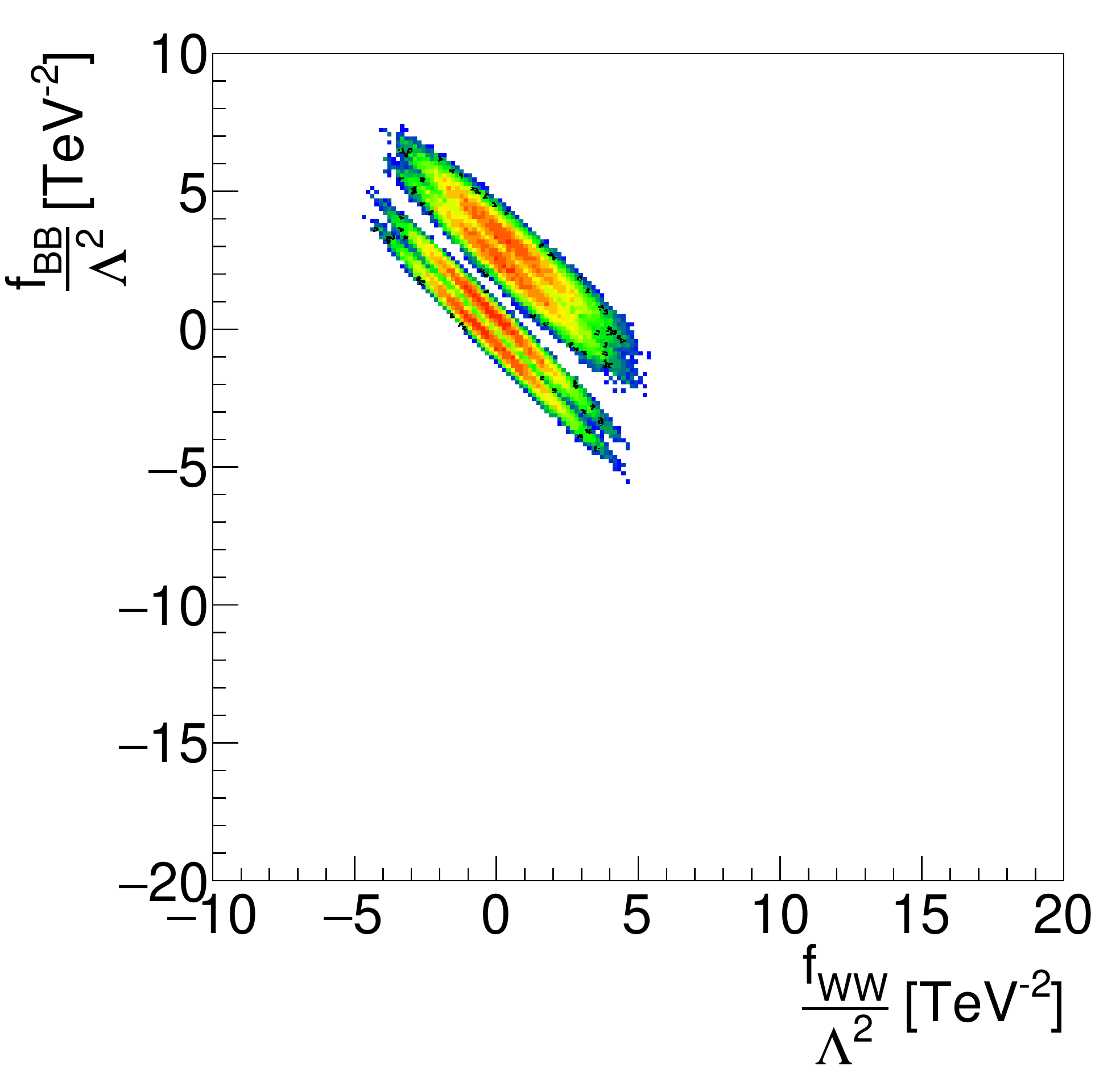}
  \hspace*{-0.2cm}
  \includegraphics[width=0.3\textwidth]{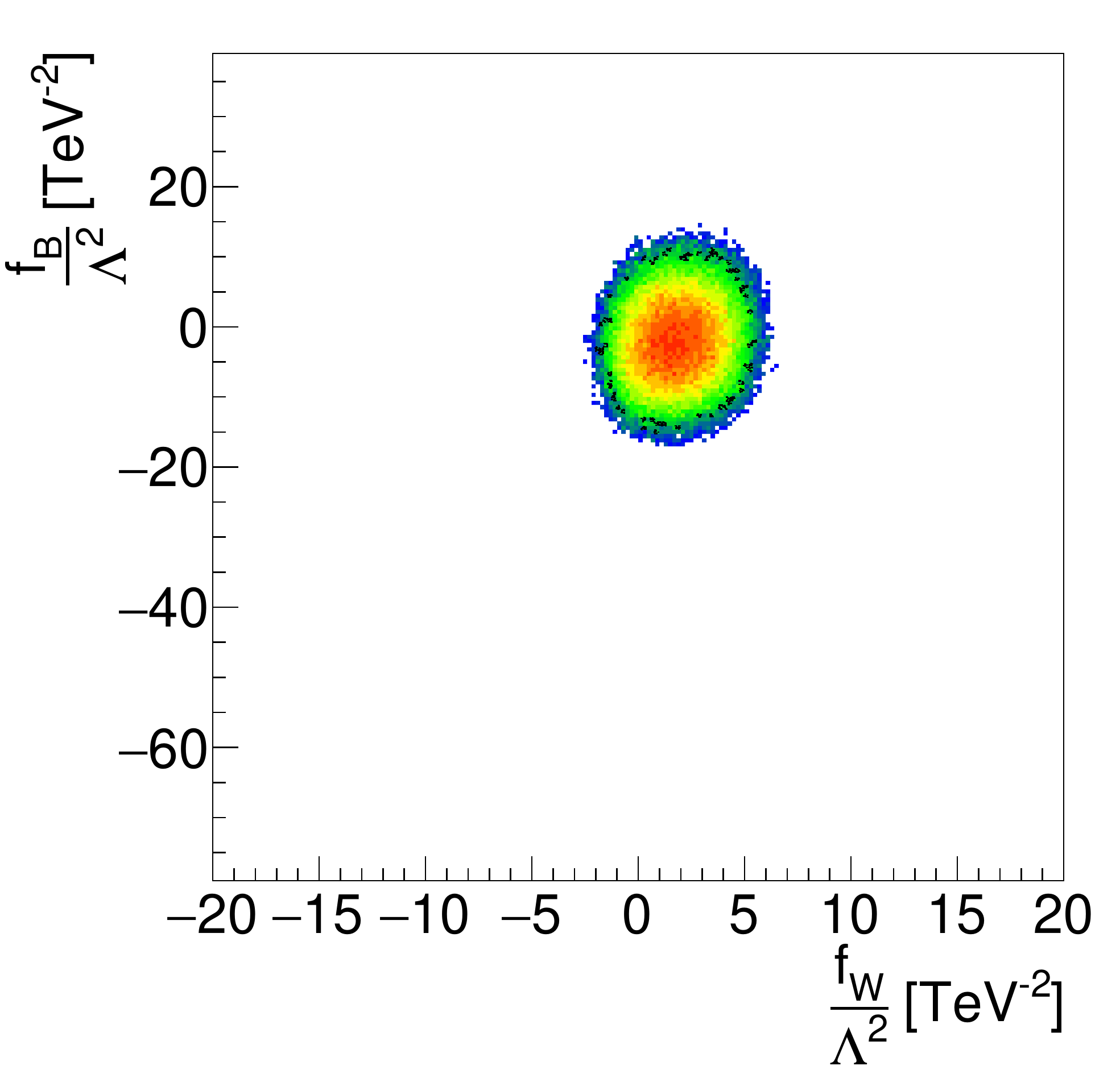}
  \hspace*{-0.2cm}
  \includegraphics[width=0.3\textwidth]{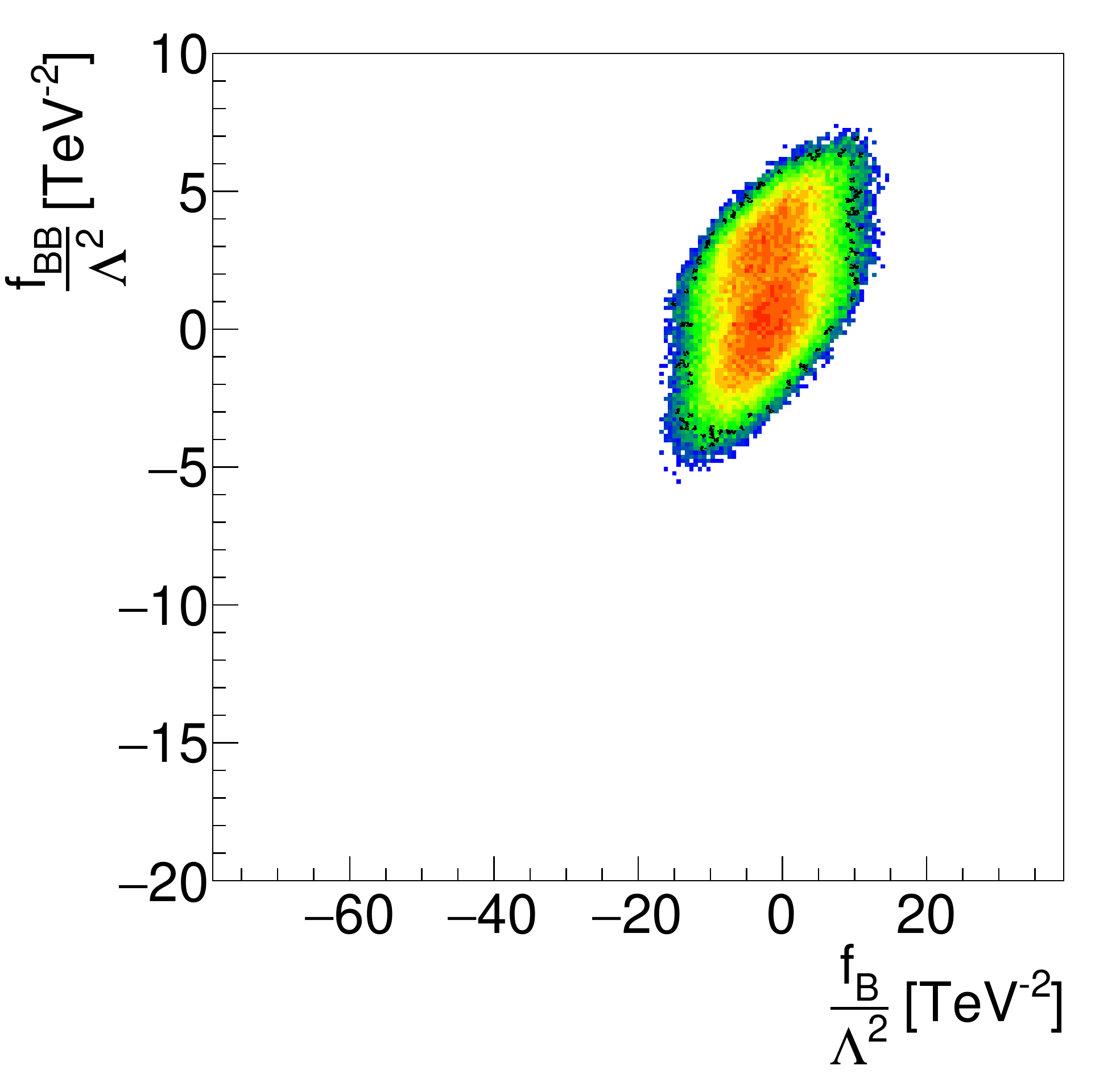}
  \raisebox{3pt}{\includegraphics[width=0.051\textwidth]{barcode.pdf}}\\[1ex]
  \caption{Correlated profile likelihood for sets of two Wilson
    coefficients. In the first row we include only LHC Run~I Higgs
    data, including kinematic distributions, as shown in Fig.~11 of
    Ref.~\cite{legacy}. In the second row we add the Run~I di-boson
    results probing anomalous TGV interactions (as well as the
    corresponding LEP results). The black points indicate
    $-2\log L=5.99$.  The corresponding one-dimensional profile
    likelihoods can be found in Fig.~\ref{fig:errorbars}.}
\label{fig:higgs2d}
\end{figure} 

In the final step of our effective field theory analysis, we have to
combine the LHC Run~I results on TGVs and Higgs couplings. The main
reason is that $\ope_{W}$ and $\ope_{B}$ contribute to anomalous Higgs
interactions and the triple gauge boson interactions at the same
time~\cite{Hagiwara:1993qt,barca,barca2}. Consequently, a study of
the underlying Wilson coefficients should include both sets of
experimental analyses. Furthermore, the combination of the two can be
used to test the nature of the electroweak symmetry breaking
mechanism~\cite{Brivio:2013pma}.\medskip

For the sake of comparison we start with a brief summary of the global
analysis of the LHC Run~I Higgs data presented in Ref.~\cite{legacy},
where constraints on the dimension-six Wilson coefficients in
Eq.~\eqref{eq:ourlag} are derived from Higgs measurements alone.  That
data consists of 159 observables for event rates, plus 14 additional
measurements related to kinematics. This kinematic information is
crucial to disentangle the strongly correlated effects of non--SM
Lorentz structures generated by $\ope_{WW}$, $\ope_{BB}$, $\ope_{W}$,
and $\ope_{B}$. This way, the kinematic distributions significantly
improve the global Higgs fit. \medskip

As an illustration, we show three of the relevant two-dimensional
profile likelihoods from the pure Higgs analysis including kinematic
distributions in the first row of Fig.~\ref{fig:higgs2d}. In the
upper-left panel we see a strong correlation between $\ope_{WW}$ and
$\ope_{BB}$, even after including the kinematic distributions. This is
due to both operators contributing to the decay rate
$H\gamma\gamma$. Without kinematic information the wide pattern in the
upper left part simply extends to the lower right
part~\cite{legacy}. The improvement in the region of large positive
(negative) $f_{WW}$ ($f_{BB}$) appears because both operators
contribute to the $HWW$ and $HZZ$ vertices, to which the kinematic
distributions are sensitive. In the upper-center panel we show the
correlations between $\ope_{W}$ and $\ope_{B}$. While the kinematic
distributions significantly improve the situation, a secondary region
still remains for negative $f_B$. Finally, in the upper-right panel we
show the $\ope_{B}$ vs $\ope_{BB}$ plane. Again, the kinematic
information largely removes the strong correlations for negative
$\ope_B$ and $\ope_{BB}$ values.\medskip

In the lower panels of Fig.~\ref{fig:higgs2d} we depict the same
two-dimensional profile likelihoods once we include the di-boson TGV
measurements from LHC Run~I; although LEP limits hardly have any numerical
effect, they are included as well. We construct the global likelihood accounting for the
correlations in systematic uncertainties between the Higgs observables
and the TGV observables. This can be easily achieved in the
\textsc{SFitter} framework described in Sec.~\ref{sec:gauge} and
Ref.~\cite{legacy}. The systematic experimental uncertainties are
assumed to be correlated for observables in ATLAS and in CMS, but
uncorrelated between the two experiments. \medskip

For all three panels the effect of the TGV measurements is
remarkable. The combination of Higgs and TGV results clearly deliver
stronger limits than either of the two analyses independently.  The
secondary solution in $f_B$ has vanished altogether, the precision on
$f_W$ has improved, negative values of $f_{BB}$ are excluded through
correlations with $f_B$, and in the correlation of $f_{BB}$ and
$f_{WW}$ we can clearly see two different regions corresponding to
sign changes in the $H\gamma\gamma$ coupling.\medskip

\begin{figure}[t]
\includegraphics[width=0.6\textwidth]{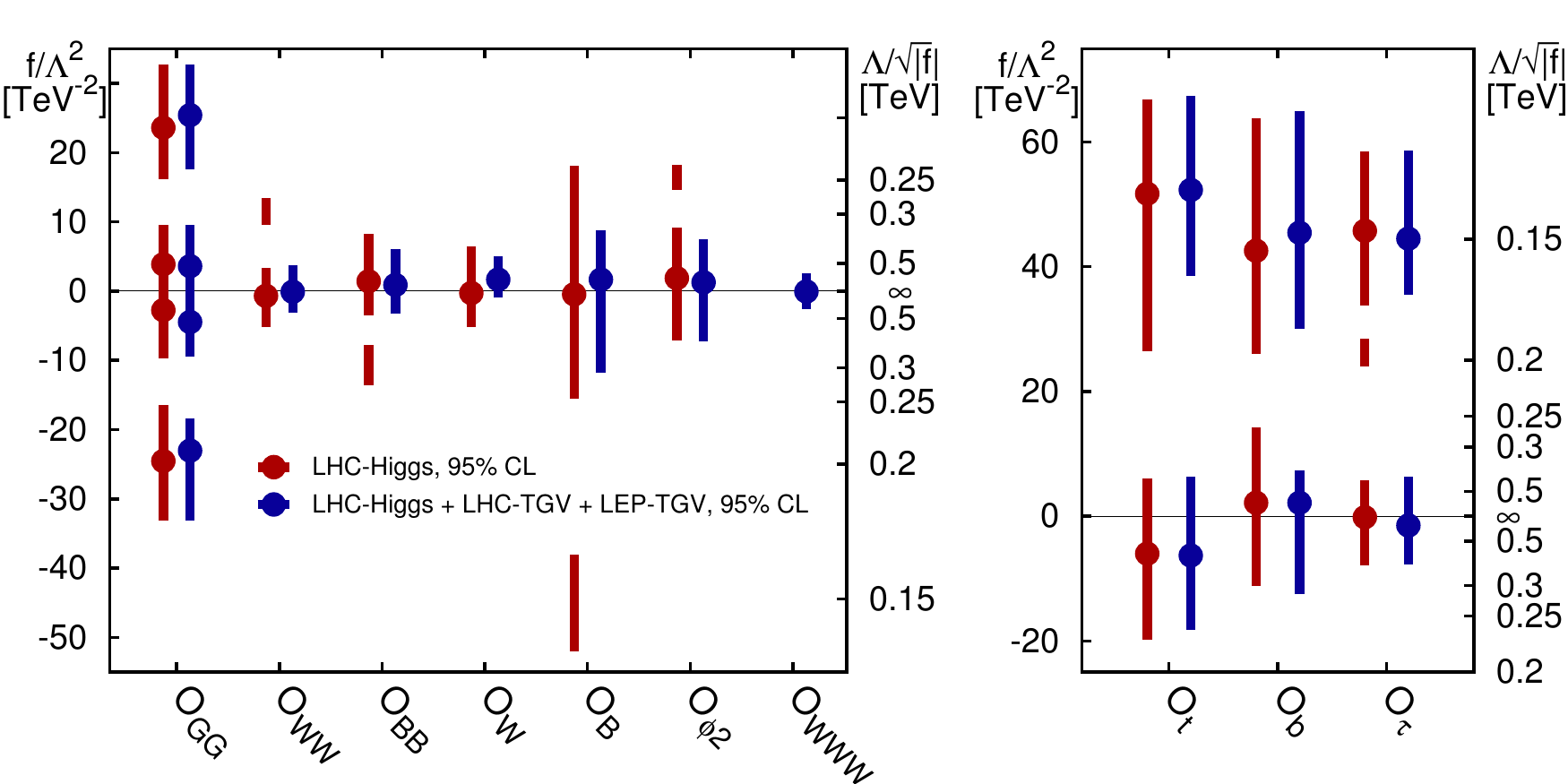} \\
\includegraphics[width=0.6\textwidth]{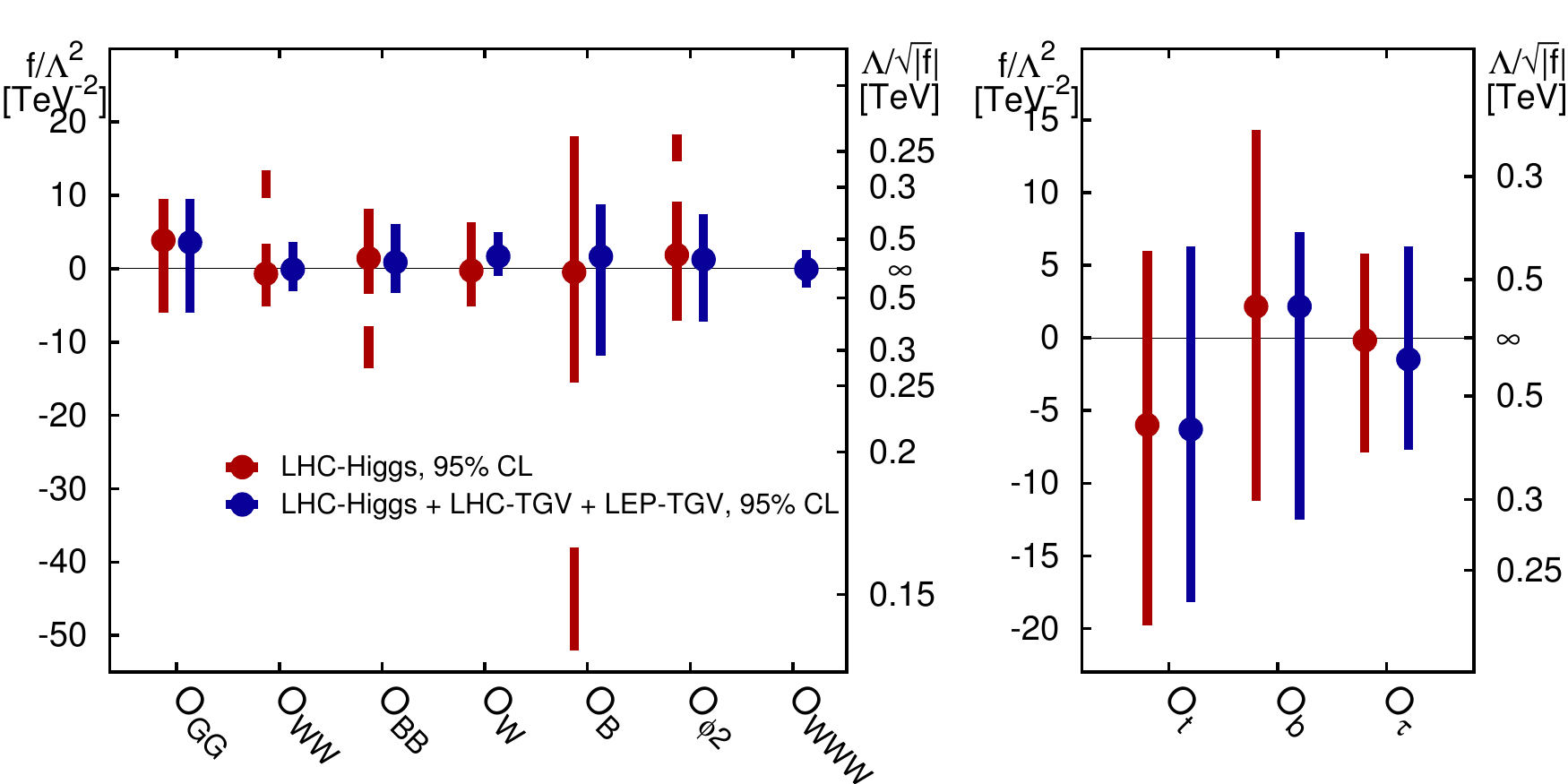}
\caption{Allowed 95\% CL ranges for individual Wilson coefficients
  $f_x/\Lambda^2$ from a one-dimensional profile likelihood. We show
  results from Run~I Higgs observables only (red bars) and for a
  combined Higgs plus TGV analysis (blue). For the upper panels we
  allow for sign changes in the individual Yukawa couplings, while in
  the lower panel we fix their signs to the Standard Model one.}
\label{fig:errorbars}
\end{figure}

In Table~\ref{tab:errorbars} and Fig.~\ref{fig:errorbars} we show the
limits on individual Wilson coefficients for each of the
dimension--six operators included in the analysis,
Eq.~\eqref{eq:ourlag}. In the upper panels of Fig.~\ref{fig:errorbars}
and in the table we clearly see secondary solutions due to sign flips
in the individual Yukawa and $Hgg$ couplings. In the lower panels of
Fig.~\ref{fig:errorbars} we show only the solutions for parameter
space with SM signs of the Yukawa couplings, and focusing on the
$f_{GG}$ containing the SM point, extending our set of simplifications
discussed in Sec.~\ref{sec:intro_th}. In both cases we see that the
limits including di-boson channels are significantly improved. This
improvement is driven by the highest sensitivity we have derived on
$f_B$ and $f_W$, which feeds through to the remaining operators
because of the existing correlations.
Including the di-boson data removes all secondary
solutions from non-trivial parameter correlations or strong
non-Gaussian effects. The additional Wilson coefficient $f_{WWW}$ is
among the best-measured dimension--six modification in the gauge--Higgs
sector studied here. \medskip

\begin{table}[t!]
\begin{tabular}{l|rc|rc}
\hline
$f_x/\Lambda^2 [\tev^{-2}]$ 
&  \multicolumn{2}{c|}{LHC--Higgs} &  \multicolumn{2}{c}{LHC--Higgs + LHC--TGV + LEP--TGV} \\
& Best fit & 95\% CL interval & Best fit & 95\% CL interval \\
\hline
$f_{GG}$ & $-24.5$ & $(-33.2, 16.4)$ & $-4.5$  & $(-32.5, -18.4)$  \\
        & $-2.8$ & $(-9.7, 9.5)$ & $-23.0$  & $(-9.5, 9.5)$ \\
        & 3.9 & $(16.2, 32.7)$ & 3.6  & $(17.6, 32.5)$ \\
        & 23.6 & & 25.4  & \\ \hline
$f_{WW}$ & $-0.7$ & $(-5.2, 3.4)\cup(9.6, 13.4)$ & $-0.1$ & $(-3.1, 3.7)$ \\
$f_{BB}$ & 1.4 & $(-13.6, -7.8)\cup(-3.5, 8.2)$ & 0.9 & $(-3.3, 6.1)$ \\
$f_{\phi,2}$ & 1.9 & $(-7.1, 9.2)\cup(14.6, 18.3)$ & 1.3 & $(-7.2, 7.5)$ \\
$f_{W}$  & $-0.3$ & $(-5.2, 6.4)$ & 1.7 & $(-0.98, 5.0)$ \\
$f_{B}$  & $-0.5$ & $(-52, -38)\cup(-15.5, 18.1)$ & 1.7 & $(-11.8, 8.8)$ \\
$f_{WWW}$ & \multicolumn{2}{c|}{-----} & -0.06 & $(-2.6, 2.6)$ \\ \hline
$f_{b}$  & 2.2 & $(-11.2, 14.3)$ & 2.2 & $(-12.5, 7.3)$ \\
         & 42.6 & $(26, 64)$ & 45.6 & $(30, 65)$ \\
$f_{\tau}$ & 45.8 & $(-7.9, 5.8)\cup(24,28)$ & 44.5 & $(-7.7, 6.3)$ \\
          & $-0.2$ & $(34, 59)$ & $-1.5$ & $(36, 59)$ \\
$f_{t}$ & 51.8 & $(-19.8, 6.0)$ & 52.3 & $(-18.2, 6.3)$ \\
        & $-6.0$ & $(27, 67)$ & $-6.3$ & $(39, 68)$ \\
\hline
& \multicolumn{2}{c|}{$(-2 \log L)_\text{min}=98.1$, $(-2 \log L)_\text{SM}=101.9$} 
& \multicolumn{2}{c}{$(-2 \log L)_\text{min}=152.3$, $(-2 \log L)_\text{SM}=156.8$} \\ 
\hline
\end{tabular}
\caption{Best fit values and 95\% CL ranges for the Higgs analysis
  (dark red bars in Fig.~\ref{fig:errorbars}) and after combining with
  TGV results (blue bars in Fig.~\ref{fig:errorbars}).  We also show
  log-likelihood values, where $(-2 \log L)_\text{SM}$ is defined
  after profiling over the theoretical uncertainties.}
\label{tab:errorbars}
\end{table}

One caveat applies to these results the same way it applies to the
Higgs analysis alone~\cite{legacy,d6_vs_eft}. If we consider the
Lagrangian of Eq.~\eqref{eq:ourlag} to be the leading term in a
systematic effective field theory, we have to ensure that only data
probing typical momentum ranges below $\Lambda$ enters our
analysis. Otherwise, neglected dimension--eight and higher operators
might lead to large effects. This condition may not always be
fulfilled in all the bins of the kinematic distributions studied here
depending on the assumed size of the Wilson coefficients, as
illustrated in the Appendix. Therefore, we refrain from interpreting
our results in terms of an effective field theory and instead consider
Fig.~\ref{fig:errorbars} as limits on a truncated dimension-six
Lagrangian. 

\section{Summary}
\label{sec:conclu}

We have presented a final analysis of the LHC Run I measurements
related to weak boson self-interactions and Higgs decays in the
framework of an effective Lagrangian to dimension--six. The parameter
space for this analysis spans over 10 relevant Wilson coefficients
given in Eq.~\eqref{eq:ourlag}. All of them can be strongly
constrained by the combination of Higgs and di-boson data.\medskip

For triple gauge-boson data we give the first combination of all the
di-boson production channels at LHC Run~I, relevant to constrain the
three dimension--six operators contributing. The current bounds
derived in Sec.~\ref{sec:gauge} are a factor 3-6 more precise than the
corresponding LEP bounds. Since LHC Run I is sensitive to the TGVs in
a diverse set of channels, the allowed parameter ranges for the
couplings are only weakly correlated; see Eq.~\eqref{eq:cormat}. In the
future, we expect sizeable progress in particular for channels with
semi-leptonic decays of weak-boson pairs.\medskip

In Sec.~\ref{sec:higgs} we combine the Run~I di-boson data with the
Run~I Higgs measurements~\cite{legacy}. This leads to a significant
improvement compared to both individual analyses. While in the Higgs
analysis alone we are left with strong correlations between the
different Wilson coefficients --- leading to large non-Gaussian
structures in the correlated likelihood --- secondary solutions in the
combined analysis are exclusively due to the signs of the Yukawa
couplings. Furthermore, the use of the Higgs data leads to an
improvement on the determination of TGVs, specially $f_B$.
Our results shown in Fig.~\ref{fig:errorbars} clearly
indicate that di-boson data should be part of any effective Lagrangian
analysis of the Higgs sector at the LHC.
%
\bigskip

\begin{center} \textbf{Acknowledgments} \end{center}

A.B. is supported by the Heidelberg Graduate School for Fundamental Physics and the Graduiertenkolleg
Particle Physics beyond the Standard Model. O.J.P.E. is supported in
part by Conselho Nacional de Desenvolvimento Cient\'{\i}fico e
Tecnol\'ogico (CNPq) and by Funda\c{c}\~ao de Amparo \`a Pesquisa do
Estado de S\~ao Paulo (FAPESP). \mbox{M.C.G-G} acknowledges support from
USA-NSF grant PHY-13-16617, EU grant FP7 ITN INVISIBLES (Marie Curie
Actions PITN-GA-2011-289442), FP10 ITN ELUSIVES (H2020-MSCA-ITN-2015-674896),
INVISIBLES-PLUS (H2020-MSCA-RISE-2015-690575), Spanish grants 2014-SGR-104 and by
FPA2013-46570 and consolider-ingenio 2010 program CSD-2008-0037.

\clearpage

\appendix
\section{ATLAS WW analysis}
\label{sec:app}

In this appendix we describe in detail how we include the experimental
results in our \textsc{SFitter} analysis. As an example we use one of
the most sensitive channels, namely the leptonic ATLAS $WW$ analysis
based on $20.3~\ifb$ of 8~TeV data~\cite{atlas8ww}. One advantage of
this analysis is that ATLAS presents their results in terms of TGVs,
which allows us to compare their results with the ones of our
\textsc{SFitter} implementation. The other seven channels are treated
exactly in the same way.\medskip

We start by generating $WW$ events with SM couplings using
\textsc{MadGraph5}~\cite{madgraph},
\textsc{Pythia}~\cite{Sjostrand:2006za} for parton shower and
hadronization, and \textsc{Delphes}~\cite{delphes} for fast detector
simulation. We model here the ATLAS selection, which is very similar to
the analogous CMS analysis~\cite{cms8ww}. The selection procedure
requires exactly one electron and one muon of opposite charges in the
central detector and outside the transition regions,
\begin{alignat}{9}
 p_{T,\ell} &>25, 20~\gev  \qqqquad
 &|\eta_\mu|&<2.4   \qqqquad 
 &|\eta_e|&<2.47 \quad  \text{excluding} \; 1.37<|\eta_e|<1.52 \notag \\
 \Delta R_{e\mu}&>0.1 
 &m_{e\mu}&>10~\gev \; .
\end{alignat}
In addition, the summed transverse energy within a cone of
$\Delta R=0.3$ around each lepton is required to be smaller than 14\%
of $p_{T,\ell}$, and the scalar sum of the $p_T$ of the tracks within
the same cone has to stay below 10\% of $p_{T,\ell}$ for the electron
and 12\% for the muon. A third lepton is vetoed for
$p_{T,\ell}>7$~GeV, as are jets with $p_{T,j}>25$~GeV and
$|\eta_j|<4.5$. The latter removes the top pair background. A set of
requirements on missing energy related variables starts with a
requirement on $p_T^{\;\text{miss}}$, constructed as the length of
the negative 2-vectorial sum of all identified leptons and tracks not
associated with leptons~\cite{atlas8ww}. To select events with
neutrinos ATLAS requires
\begin{align}
p_{T}^{\;\text{miss}}>20~\gev 
\qquad \text{and} \qquad
\Delta\phi\left(\vec{E}_T^{\;\text{miss}},\vec{p}_T^{\;\text{miss}}\right)<0.6\; . 
\end{align}
A second missing energy variable has to fulfill
\begin{align}
 E_{T,\text{Rel}}^{\;\text{miss}}>15~\gev
 \qquad \text{with} \quad 
 E_{T,\text{Rel}}^{\;\text{miss}}
 =\begin{cases}
  E_T^{\;\text{miss}}\sin(\Delta\phi_\ell) & \text{if} \quad \Delta\phi_\ell<\pi/2 \\
  E_T^{\;\text{miss}} & \text{if} \quad \Delta\phi_\ell\geq\pi/2 \; ,
  \end{cases}
\end{align}
where $\Delta\phi_\ell$ is the azimuthal angle between the missing
transverse momentum vector and the nearest lepton.\medskip

\begin{figure}[b!]
\centering
\includegraphics[width=0.45\textwidth]{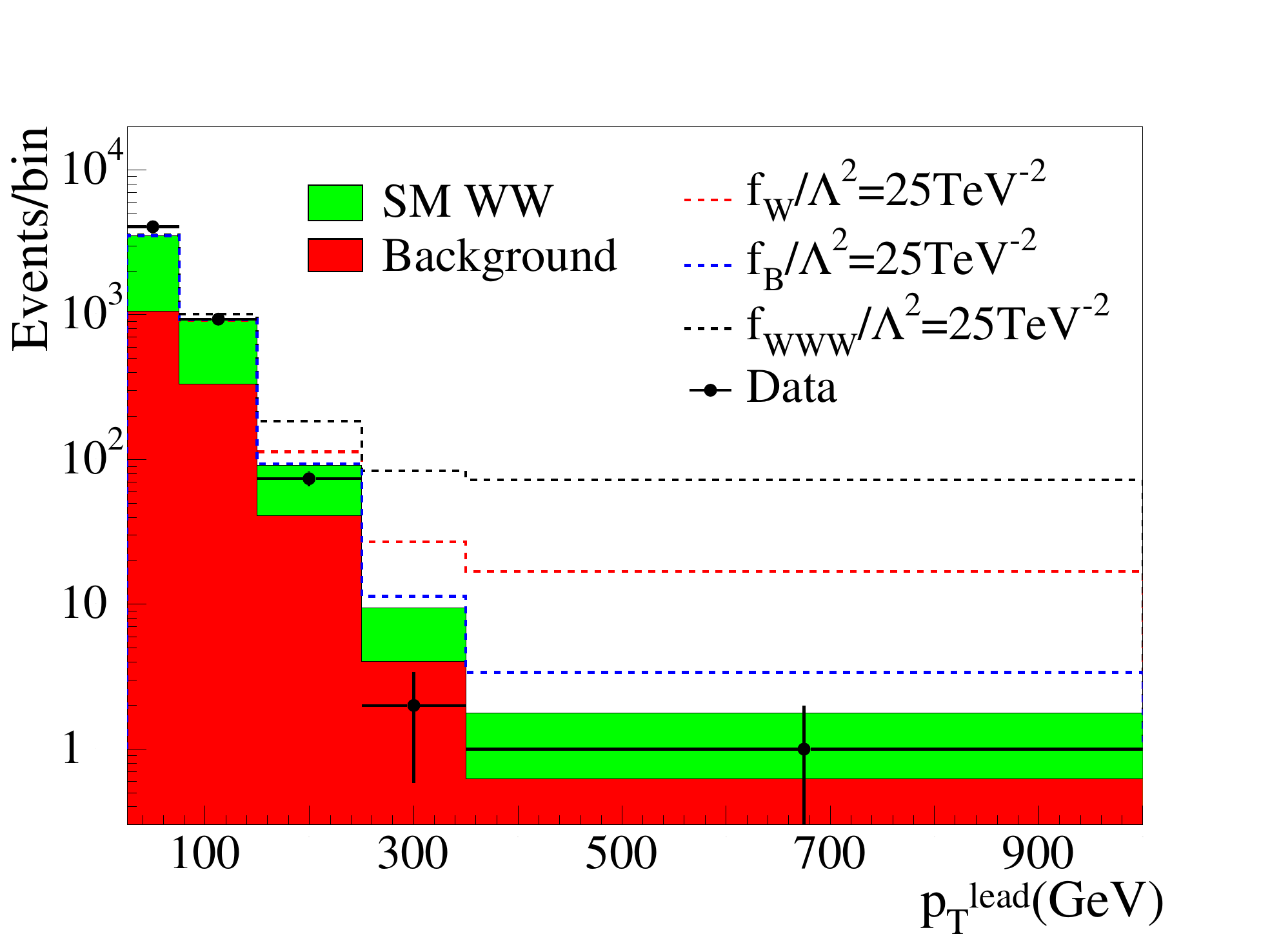}
\caption{Leading $p_{T,\ell}$ distribution for the 8~TeV ATLAS $WW$
  analysis~\cite{atlas8ww}. The red histogram shows the ATLAS
  background estimate (excluding the SM $WW$ prediction), while the
  green histogram shows the total SM prediction once $WW$ processes
  are added.  The observed events are shown as dots, with error bars
  accounting for the statistical uncertainty. The dashed lines
  indicate the effects of dimension--six Wilson coefficients.}
\label{fig:ptlead}
\end{figure}

We use the SM $WW$ events rates in the signal region to tune our event
generation, both in terms of the total rate and in the most relevant
kinematic distribution. For the latter, we identify the kinematic
distribution which is most sensitive to anomalous TGVs and which we
will later include in our \textsc{SFitter} analysis.  Of the variables
and ranges shown in the ATLAS note, the leading $p_{T,\ell}$ has the
largest potential because it tracks the momentum flow through the
anomalous vertex best~\cite{Eboli:2010qd,anke_johann}. This means that
our event generation has to reproduce Fig.~11 in
Ref.~\cite{atlas8ww}. To ensure this, we introduce a bin-by-bin
correction factor to account for differences in the selection
procedure because of detector effects as well as higher order
corrections to the cross section prediction~\cite{nloww}.  \medskip

Assuming that the same bin-by-bin correction from the SM $WW$ events
applies to the relatively small new physics effects, we generate the
leading $p_{T,\ell}$ distribution in the presence of dimension--six
operators. For this we rely on \textsc{MadGraph5} and an in-house
implementation of the operators through
\textsc{FeynRules}~\cite{Christensen:2008py}.
%
As is well known higher dimensional operators give rise to fast
  growth of the scattering amplitude with energy, eventually violating
  partial-wave unitarity~\cite{Corbett:2014ora}. Here we did not
  introduce ad-hoc form factors to dampen the scattering amplitude at
  high energies because we verified that there is no partial-wave
  unitarity violation in the different channels for the values of the
  Wilson coefficients in the 95\% CL allowed regions, except for very
  large and already ruled out values of $f_B$. \medskip

The predicted number of events for a given Wilson coefficient is the sum
of SM and new physics $WW$ events, together with the SM backgrounds
which we directly extract from the ATLAS documentation. These
backgrounds are dominated by top production, followed by $W$+jets  and
Drell-Yan events. All of them are estimated using data-driven
techniques. Only the small di-boson backgrounds are based on Monte
Carlo estimates~\cite{atlas8ww}. \medskip

In Fig.~\ref{fig:ptlead} we show the final estimates for the SM
background and the SM prediction for $WW$ production. They are in
agreement with the number of observed events. The dashed lines
illustrate the effects from individual dimension-six operators,
suggesting that we should be able to derive powerful constraints from
the ATLAS measurements.  The fact that the last bin extends to very
large transverse momenta also suggests that we have to be careful
interpreting our dimension-six analysis in terms of an effective field
theory expansion~\cite{legacy,d6_vs_eft}.\medskip

\begin{figure}[t!]
\centering
\includegraphics[width=0.3\textwidth]{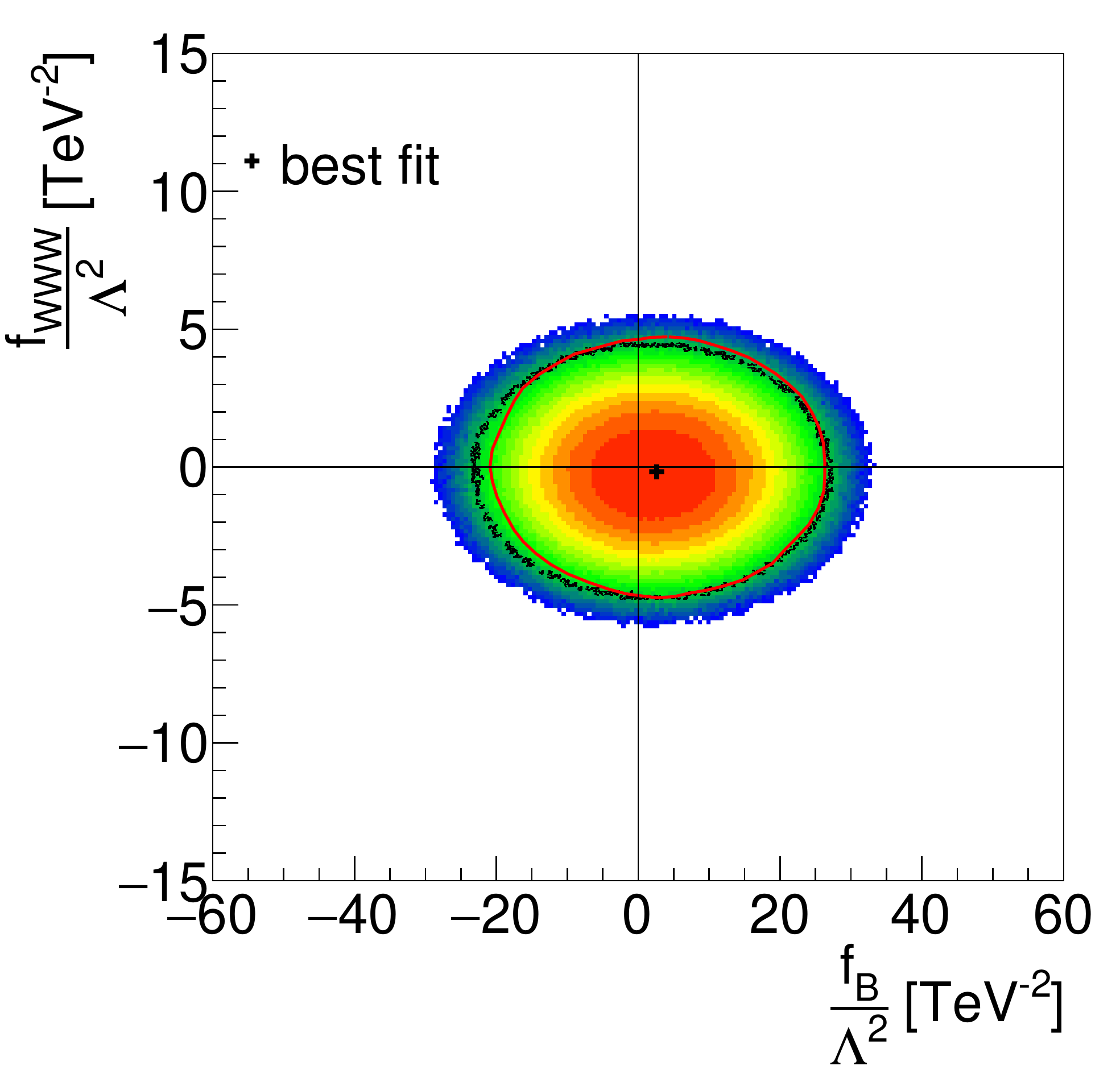}
\includegraphics[width=0.3\textwidth]{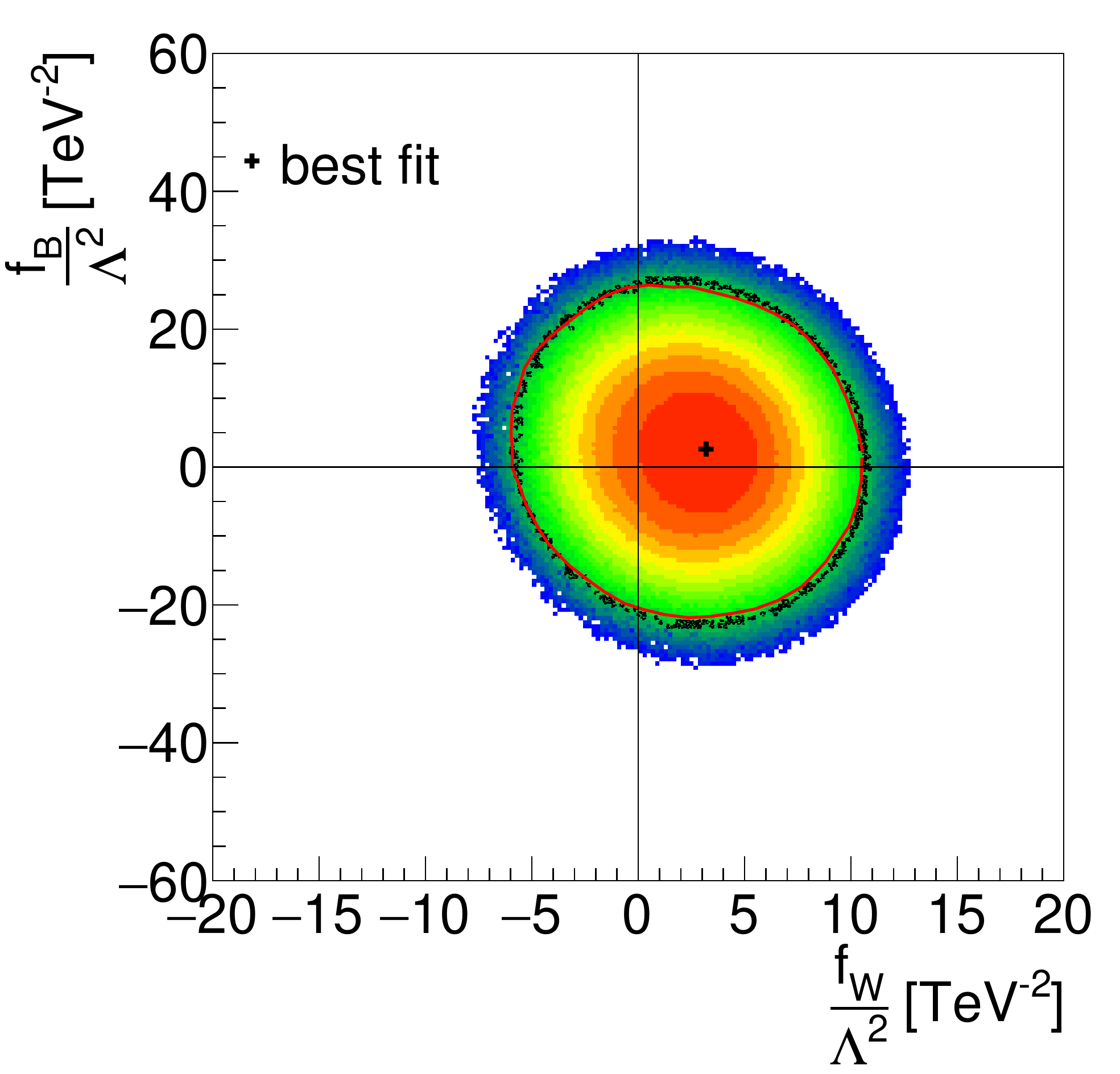}
\includegraphics[width=0.3\textwidth]{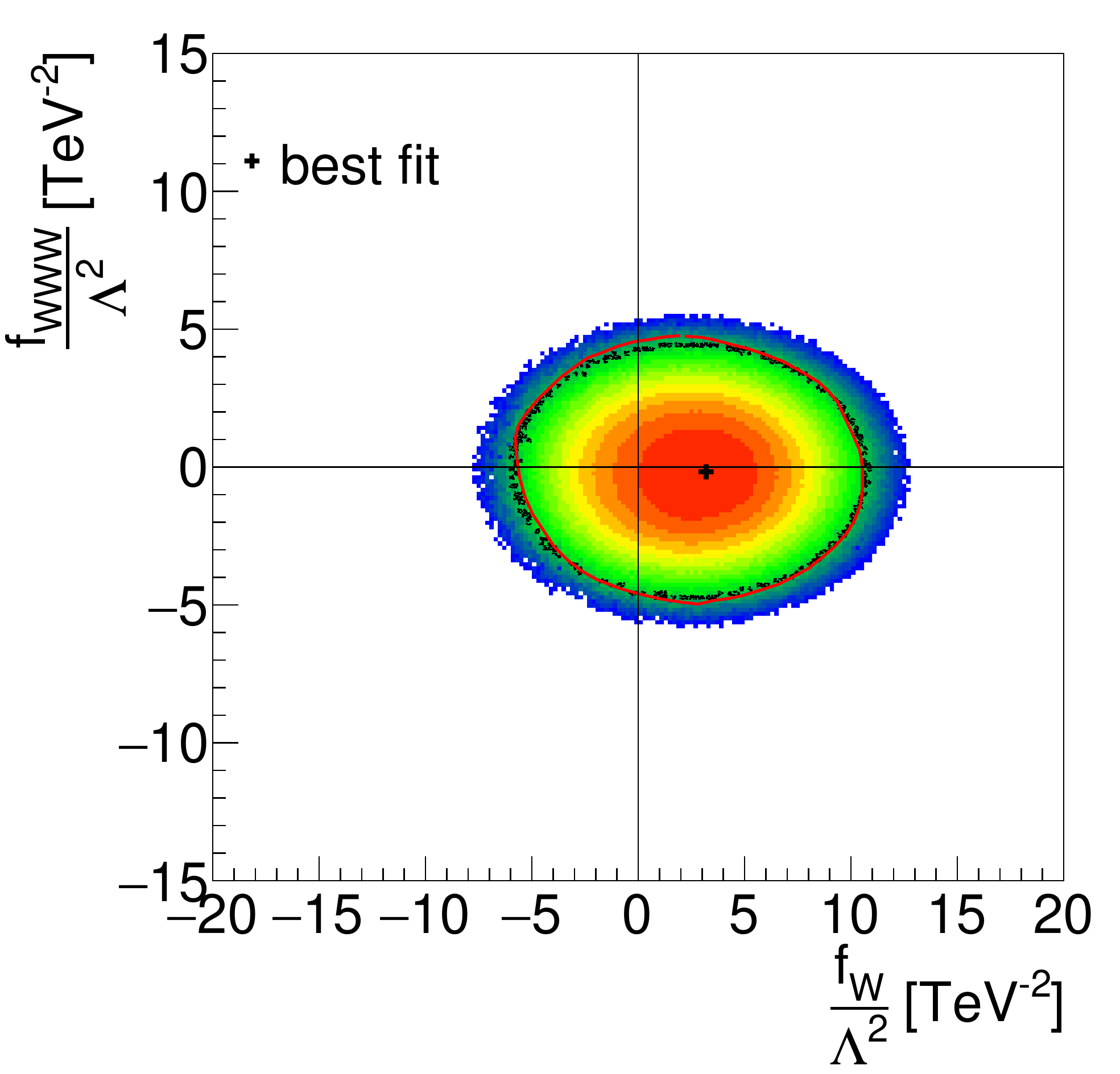}
\raisebox{3pt}{\includegraphics[width=0.051\textwidth]{barcode.pdf}}
\caption{Correlated profile likelihood for sets of two Wilson
  coefficients from the 8~TeV ATLAS $WW$
  analysis~\cite{atlas8ww}. Black dots signal $\Delta(-2\log L)=5.99$,
  while the crosses stand for the best fit point. The red solid
  contour are the 95\% CL limits from ATLAS~\cite{atlas8ww}. }
\label{fig:AT8WW}
\end{figure}

In the final step of the \textsc{Sfitter} analysis we construct a
likelihood function including a Poisson-shaped statistical uncertainty
for the observed number of events for each bin, a Poisson-shaped
statistical uncertainty for the background events, a flat theoretical
uncertainty correlating between all bins in the $p_{T,\ell}$
distribution, and a selection of the most relevant systematic
uncertainties with a Gaussian shape. These uncertainties can be seen
in the following, together with the selection of experimental
systematic uncertainties considered for the rest of the analyses.

\begin{center} \begin{tabular}{ll|rrrr}
\hline 
Channel & Exp & Luminosity & Detector eff & Lepton eff & Background rate \\ 
\hline
$WW\rightarrow \ell^+\ell^{\prime -}+\sla{E}_T\; (0j)$~\cite{atlas8ww} & ATLAS    & 2.0\% & 1.4\% & 1.4\% & 2.0\% \\
$WW\rightarrow \ell^+\ell^{(\prime) -}+\sla{E}_T\; (0j)$~\cite{cms8ww}  & CMS     & 2.6\% & 1.0\% & 3.8\% & 2.0\% \\
$WZ\rightarrow \ell^+\ell^{-}\ell^{(\prime)\pm}$~\cite{atlas8wz}         & ATLAS  & 2.8\% & 0.5\% & 1.7\% & 1.6\% \\
$WZ\rightarrow \ell^+\ell^{-}\ell^{(\prime)\pm}+\sla{E}_T$~\cite{cms78wz} & CMS   & 4.4\% & 3.1\% & 2.0\% & 2.5\% \\
$WV\rightarrow \ell^\pm jj+\sla{E}_T$~\cite{atlas7semilep}              & ATLAS & 1.8\% & 10\% & 1.1\% & 14\% \\
$WV\rightarrow \ell^\pm jj+\sla{E}_T$~\cite{cms7semilep}                & CMS   & 2.2\% & 1.0\% & 2.0\% & -- \\
$WZ\rightarrow \ell^+\ell^{-}\ell^{(\prime)\pm}+\sla{E}_T$~\cite{atlas7wz}& ATLAS & 1.8\% & 0.5\% & 1.9\% & -- \\
$WZ\rightarrow \ell^+\ell^{-}\ell^{(\prime)\pm}+\sla{E}_T$~\cite{cms78wz} & CMS   & 2.2\% & 3.8\% & 2.4\% & 5.5\% \\
\hline
\end{tabular} \end{center}

\noindent
For the cases where we quote no numbers we assume that those systematic
uncertainties are well below the statistical and theoretical
uncertainties. For the pure TGV analysis we construct Markov chains
to probe the three-dimensional parameter space spanned by $f_W$, $f_B$
and $f_{WWW}$.  Based on these chains we determine the part of the
parameter space allowed at a given CL.\medskip

In Fig.~\ref{fig:AT8WW} we show the three two-dimensional profile
likelihoods for the three relevant Wilson coefficients. We find the
best-fit point for a mildly positive value of $f_W/\Lambda^2$, driven
by a small deficit of events in the tail of the leading $p_{T,\ell}$
distribution shown in Fig.~\ref{fig:ptlead}. The SM gives
$\chi^2 \approx - 2 \log L = 6.6$, defined after profiling over the
theoretical uncertainties, and is perfectly compatible with the best
fit point at $\chi^2 \approx -2 \log L= 6.0$. We have checked that
none of these results change if we replace the profile likelihood by a
slice of parameter space setting the third Wilson coefficient to
zero. \medskip

The black dots in Fig.~\ref{fig:AT8WW} indicate our 95\% CL contour
and allow us to compare with the red line, that illustrates the 95\%
CL region from the anomalous TGV analysis by
ATLAS~\cite{atlas8ww}. Both are in excellent agreement with each
other, indicating that our approximations concerning detector effects
or higher order corrections are more than sufficient given the current
reached precision of the analysis. \medskip

We follow a similar procedure for all eight di-boson channels.  Among
those the 8~TeV CMS $WW$ analysis~\cite{cms8ww} and the semi-leptonic
7~TeV ATLAS $WV$ analysis~\cite{atlas7semilep} quote limits on
dimension--six operators from the measurement of anomalous TGVs in the
framework of Eq.~\eqref{eq:lep}. In both cases we find a similar level
of agreement.


\end{document}